\documentclass[11pt,letterpaper,twoside
]{article}

\usepackage[T1]{fontenc}
\usepackage[latin1]{inputenc}
\usepackage[UKenglish]{babel}

\usepackage[pagewise]{lineno}
\usepackage{blindtext}
\usepackage{fancyhdr}

\usepackage[charter]{mathdesign}

\usepackage{amsmath}
\usepackage{amscd}
\usepackage{amsbsy}
\usepackage{amsthm}
\usepackage{nicefrac}
\usepackage{relsize}
\usepackage{accents}
\usepackage{xstring}
\usepackage{titlesec}
\usepackage{lastpage}
\usepackage[inline]%
{enumitem}
\usepackage[
longnamesfirst]{natbib}
\usepackage{color}
\usepackage[top=1in,bottom=1in,left=1in,right=1in]{geometry}
\usepackage{tikz}
\usepackage[hang]{footmisc}
\usepackage[
bookmarks=false,colorlinks=true,linkcolor=DCUBlue2,citecolor=DCUBlue2,filecolor=DCUBlue2,urlcolor=DCUBlue2,%
pdftitle={On Utility Maximisation Under Model Uncertainty in Discrete-Time Markets}
,pdfauthor={Miklos Rasonyi and Andrea Meireles-Rodrigues},%
pdfsubject={We study the problem of maximising terminal utility for an agent facing model uncertainty, in a frictionless discrete-time market with one safe asset and finitely many risky assets. We show that an optimal investment strategy exists if the utility function, defined either on the positive real line or on the whole real line, is bounded from above. We further find that the boundedness assumption can be dropped provided that we impose suitable integrability conditions, related to some strengthened form of no-arbitrage. These results are obtained in an alternative framework for model uncertainty, where all possible dynamics of the stock prices are represented by a collection of stochastic processes on the same filtered probability space, rather than by a family of probability measures.},
pdfkeywords={Discrete\ time;\ } {model\ uncertainty;\ } {optimal\ portfolio;\ } {utility\ maximisation.}]{hyperref}
\hypersetup{pdfstartview={XYZ null null 1.00}}
\usepackage{thmtools}
\usepackage{txfonts}
\usepackage[backgroundcolor=white,linecolor=DCUPink,bordercolor=DCUPink,textsize=footnotesize,textwidth=2cm%
]{todonotes}
\usepackage[compress]{cleveref}
\usepackage{lipsum}
\usepackage{float}
\usepackage
{placeins}
\usepackage[font={normal,it},
labelfont=normal]{caption}
\usepackage{subcaption}
\usepackage{comment}
\usepackage{epstopdf}
\usepackage{datetime}

\crefname{sec}{section}{sections}
\crefname{subsec}{subsection}{subsections}
\crefname{def}{definition}{definitions}
\crefname{th}{theorem}{theorems}
\crefname{lem}{lemma}{lemmata}
\crefname{prop}{proposition}{propositions}
\crefname{cor}{corollary}{corollaries}
\crefname{ex}{example}{examples}
\crefname{as}{assumption}{assumptions}
\crefname{obs}{remark}{remarks}
\crefname{app}{appendix}{appendices}

\definecolor{UniBlue}{RGB}{0,50,95}
\definecolor{DCUBlue}{cmyk}{1,0.64,0,0.6}
\definecolor{DCUBlue2}
{RGB}{0,35,255}
\definecolor{DCURed}{RGB}{193,0,67}
\definecolor{DCUGreen}{cmyk}{0.59,0,1,0.07}
\definecolor{MyGreen}{RGB}{100,167,11}
\definecolor{MyGreen2}{RGB}{170,210,125}
\definecolor{DCUPink}{cmyk}{0,1,0.63,0.12}
\definecolor{DCUOchre}{cmyk}{0,0.35,0.85,0}
\definecolor{MyOchreD}{RGB}{215,129,0}
\definecolor{MyOchreL}{RGB}{255,235,204}
\definecolor{MyGray}{RGB}{166,166,166}
\definecolor{MyBlue}{RGB}{166,187,179}

\bibpunct{[}{]}{;}{n}{}{,}

\makeatletter
\let\cref@old@resetby\cref@resetby%
\def\cref@resetby#1#2{
  \let#2\relax%
  \ifnum\pdfstrcmp{#1}{enumii}=\z@
    \def#2{enumi}%
  \fi%
  \ifnum\pdfstrcmp{#1}{enumiii}=\z@
    \def#2{enumii}%
  \fi%
  \ifnum\pdfstrcmp{#1}{enumiv}=\z@
    \def#2{enumiii}%
  \fi%
 \ifnum\pdfstrcmp{#1}{enumv}=\z@
    \def#2{enumiv}%
  \fi%
 \ifx#2\relax%
    \cref@old@resetby{#1}{#2}
 \fi}%
\makeatother

\setlist[enumerate,1]{
  leftmargin=*,
}
\setlist[enumerate,2]{
	wide=0pt
}

\newcommand{\ADD}%
[3][]{%
	\IfEqCase{#2}{%
		{1}{\textcolor{%
		black
		}{#3}}%
    {0}{#1}%
	}[]%
}%

\makeatletter
\newcommand{\labeltarget}[1]{\Hy@raisedlink{\hypertarget{#1}{}}}
\makeatother

\newcommand{\pprangestart}[1]{\label{#1START}}
\newcommand{\pprangeend}[1]{\label{#1END}}
\newcommand{\pprange}[1]{
	\ifthenelse{\equal{\pageref{#1START}}{\pageref{#1END}}}{p.\,\pageref{#1START}}{pp.\,\pageref{#1START}--\pageref{#1END}}}

\newcommand{\bbOne}{1\hspace*{-0.8ex}1}

\newcommand{\Lspace}[4][]{L^{#2}_{#3}\!\left({#4};\mathbb{R}^{#1}\right)}

\newcommand{\norm}[2][\mathbb{R}^{d}]{\left\Vert{#2}\right\Vert_{#1}}
\newcommand{\innerp}[3][\mathbb{R}^{d}]{\left\langle{#2},{#3}\right\rangle_{#1}}

\newcommand{\iw}{w_{0}}

\newcommand{\adm}[1]{\mathscr{A}\!\left({#1}\right)}
\newcommand{\admR}[1]{\bar{\mathscr{A}}\!\left({#1}\right)}
\newcommand{\admE}[1]{\tilde{\mathscr{A}}\!\left({#1}\right)}

\newcommand{\rcd}[3][]{\mathbb{P}^{\left. #2 \right|\mathscr{#3}_{#1}}}

\newcommand{\seq}[2][n]{\left\{{#2}\right\}_{{#1}\in\mathbb{N}}}

\newcommand{\vect}[2][d]{\boldsymbol{#2_{#1}}}

\newcommand{\safe}{%
\boldsymbol{0}%
}

\newcommand{\NAset}{%
\mathscr{S}^{\ast}%
}

\newcommand{\opt}{%
\phi^{\ast}%
}

\newcommand{\spd}{\zeta}

\newcommand{\ubar}[1]{\underaccent{\bar}{#1}}

\newcommand{\pow}[1]{%
2^{#1}%
}

\newcommand{\FL}[1]{%
V%
}

\DeclareMathOperator{\conv}{conv}
\DeclareMathOperator{\dom}{dom}
\DeclareMathOperator{\inter}{int}
\DeclareMathOperator{\proj}{proj}
\DeclareMathOperator{\Law}{Law}
\DeclareMathOperator*{\esssup}{ess\,sup}
\DeclareMathOperator*{\essinf}{ess\,inf}

\numberwithin{equation}{section} 

\newcommand{\addperiod}[1]{{#1}.}

\titleformat{\section}{\normalfont\Large\bfseries}{\thesection.}{0.55em}{
}[]
\titleformat{\subsection}{\normalfont\large\bfseries}{\thesubsection.}{0.55em}{}
\titleformat{\subsubsection}{\normalfont\normalsize\bfseries}{\thesubsubsection.}{0.55em}{}
\titleformat{\paragraph}[runin]{\normalfont\normalsize\itshape}{\theparagraph.}{0.55em}{\addperiod%
}

\declaretheoremstyle[spaceabove=0.5\topsep,spacebelow=0.5\topsep,headfont=\scshape,notefont=\scshape, notebraces={(}{)},headpunct={.},bodyfont=\normalfont\itshape,headindent=0pt,postheadspace={ }]{mytheorem}
\declaretheorem[name=Theorem,numberwithin=section,style=mytheorem]{theorem}
\declaretheorem[name=Proposition,sibling=theorem,style=mytheorem]{proposition}
\declaretheorem[name=Lemma,sibling=theorem,style=mytheorem]{lemma}

\declaretheorem[name=Assumption,sibling=theorem,style=mytheorem]{assumption}

\declaretheoremstyle[spaceabove=0.5\topsep,spacebelow=0.5\topsep,headfont=\scshape,notefont=\scshape, notebraces={(}{)},headpunct={.},bodyfont=\normalfont,headindent=0pt,postheadspace={ }]{mydefinition}
\declaretheorem[name=Definition,sibling=theorem,style=mydefinition]{definition}

\declaretheoremstyle[spaceabove=0.5\topsep,spacebelow=0.5\topsep,headfont=\normalfont\scshape,notefont=\normalfont\scshape, notebraces={\normalfont\scshape(}{\normalfont\scshape)},headpunct={\normalfont\scshape.},bodyfont=\normalfont,headindent=0pt,postheadspace={ }]{myremark}
\declaretheorem[name=Remark,sibling=theorem,style=myremark
]{remark}

\declaretheoremstyle[spaceabove=0.5\topsep,spacebelow=0.5\topsep,headfont=\scshape,notefont=\scshape, notebraces={(}{)},headpunct={.},bodyfont=\normalfont,headindent=0pt,postheadspace={ }]{myexample}
\declaretheorem[name=Example,sibling=theorem,style=myexample]{example}

\def\shorttitle{\textsc{On Utility Maximisation under Model Uncertainty in Discrete-Time Markets%
}}

\title{\textbf{%
	\uppercase{\large%
	On Utility Maximisation under Model Uncertainty\\in Discrete-Time Markets%
	}%
}%
\thanks{R\'{a}sonyi is supported by the ``Lend\"{u}let'' grant LP2015-6 of the Hungarian Academy of Sciences and by the NKFIH (National Research, Development and Innovation Office, Hungary) grant KH~126505. Part of this research was carried out while Meireles-Rodrigues was %
affiliated with the School of Mathematical Sciences, Dublin City University, Ireland; and also while %
visiting the Alfr\'{e}d R\'{e}nyi Institute of Mathematics, whose hospitality and support are gratefully acknowledged. %
The authors wish to thank an anonymous referee for their perceptive reading and valuable comments, which have led to an improved version of this manuscript. %
They also thank Laurence Carassus for drawing their attention to certain closely related results in \citet{bc19}.%
}%
}

\author{\textsc{\normalsize\stepcounter{footnote}
Mikl\'{o}s R\'{a}sonyi}\footnote{Alfr\'{e}d R\'{e}nyi Institute of Mathematics, Re\'{a}ltanoda utca 13-15, 1053 Budapest, Hungary, email~\texttt{\href{mailto:rasonyi@renyi.hu}{rasonyi@renyi.hu}}
}\and%
\textsc{\normalsize Andrea 
Meireles-Rodrigues}\footnote{Department of Mathematics, University of York, Heslington, York, YO10~5DD, United Kingdom, %
email~\texttt{\href{mailto:andrea.meirelesrodrigues@york.ac.uk}{andrea.meirelesrodrigues@york.ac.uk}}%
}}

\newcommand{\dayend}[1]{%
	\ifthenelse{\equal{\THEDAY}{1}\OR\equal{\THEDAY}{21}\OR\equal{\THEDAY}{31}}{st}{%
		\ifthenelse{\equal{\THEDAY}{2}\OR\equal{\THEDAY}{22}}{nd}{%
			\ifthenelse{\equal{\THEDAY}{3}\OR\equal{\THEDAY}{23}}{rd}{th}}}}

\newdateformat{dateUKenglish}{\THEDAY\dayend{\THEDAY}~\monthname[\THEMONTH]~\THEYEAR}

\date%
{\normalsize \today}

\pdfoutput=1

\begin{document}


\flushbottom

\fancypagestyle{plain}{%
  \fancyhf{} 
  \renewcommand{\headrulewidth}{0pt} 
  \renewcommand{\footrulewidth}{0pt}
  \cfoot{\footnotesize \scshape{\thepage\ of {\hypersetup{linkcolor=black}\pageref{LastPage}}}}
}

\pagestyle{fancy}%
  \fancyhf{}
	\renewcommand{\headrulewidth}{0pt}
	\renewcommand{\footrulewidth}{0pt}
	\fancyhead[EL]{\footnotesize \scshape{\thepage
	}}
	\fancyhead[EC]{\footnotesize \scshape{M.\,R\'{a}sonyi and A.\,Meireles-Rodrigues}}
  \fancyhead[OR]{\footnotesize
	\thepage}
	\fancyhead[OC]{\footnotesize \shorttitle}


\thispagestyle{plain}

\maketitle

\renewcommand{\abstractname}{\normalfont\scshape Abstract}

\begin{abstract}
	\noindent%
	We study the problem of maximising terminal utility for an agent facing model uncertainty, in a frictionless discrete-time market	with one safe asset and finitely many risky assets. %
	We show that an optimal investment strategy exists if the utility function, defined either on the positive real line or on the whole real line, is bounded from above. %
	We further find that the boundedness assumption can be dropped provided that we impose suitable integrability conditions, related to some strengthened form of no-arbitrage. %
	These results are obtained in an alternative framework for model uncertainty, where all possible dynamics of the stock prices are represented by a collection of stochastic processes on the same filtered probability space, rather than by a family of probability measures.

	\bigskip
	\noindent
	{\footnotesize{
	\scshape%
	{JEL Classification:}} G11
	.}
	
	\smallskip
	\noindent
	{\footnotesize{
	\scshape%
	{AMS Mathematics Subject Classification (2010):}} 
	49K35, 
	91B16, 
	91G10, 
	93E20
	.%
	}
	
	\smallskip
	\noindent
	{\footnotesize{
	\scshape%
	{Keywords:}} discrete time; %
	model uncertainty; %
	optimal portfolio; %
	utility maximisation
	.%
	}
\end{abstract}

\clearpage


\section{Introduction}\label[sec]{sec:Intro}

Tackling model uncertainty has become a popular research area of mathematical finance%
, particularly over the past decade, in great part due to its providing a more realistic description of financial markets and behaviours. %

The classical paradigm in the literature has been that economic agents have an objective and unequivocal \emph{a priori} knowledge of the probability laws governing the evolution of markets, which is typically modelled by fixing a filtered probability space that supports an adapted stochastic process describing the prices of the underlying assets (representative papers on expected utility maximisation within the classical ``model certainty'' theory include, but are far from limited to, 
\citet{merton71
,pliska1986
,karatzasls87
,coxhuang89
,karatzaslsx91
,cvitanickaratzas1992
,zari94
,kramkov99
})%
. %
In practice, however, there is often insufficient available information to build completely accurate models. %

The dominant framework to model this uncertainty or ambiguity is to consider a non-empty family of priors $\mathscr{P}$ representing all the probability measures that are deemed possible on some canonical space where questions of hedging, pricing, and optimal investment are addressed. The presence of multiple, possibly mutually singular probability measures expresses, in an intellectually satisfying way, that we are looking for solutions which work whatever the true model specification is (in the given class of models). The associated mathematical difficulties are tantalising, just like the connection to well-established theories such as that of optimal transport (we mention the work of \citet{beiglbock2013,dolinskysoner2014}, among others).

Looking more closely at the optimal investment problem under uncertainty, the seminal paper of \citet{gilboaschmeidler1989} provides an extension to robust preferences of the axiomatic characterisation proposed by \citet{neumann53} for standard expected utility without uncertainty. %
A significant body of literature has studied this problem under the assumption that the collection $\mathscr{P}$ is dominated, i.e.\ there is a reference measure with respect to which all priors are absolutely continuous (the reader may consult e.g.\ the references cited in \citet{fsw2009,bck2019}). If, on the contrary, the set of scenario measures is non-dominated, the existence of a robust utility maximiser in discrete time has been established by: \citet{nutz2016}, for a bounded above utility function defined on the positive real line; \citet{neufeld-sikic2018}, also in the case of a utility function that is bounded from above, but in a market with frictions; \citet{bartl2017}, where a dual representation is derived for exponential utility; \citet{blanchard-carassus2017}, whose work appears to be the first treating the case of a general unbounded utility on the positive half-line; \citet{bck2019}, who rely on the axioms of ZFC set theory together with Martin's axiom (a cardinality principle weaker than the continuum hypothesis) to solve the problem for a bounded above utility on the whole real line. %

This article addresses the utility maximisation problem when uncertainty is present in an alternative framework, which we regard as no less adequate and allows to prove results that are currently unavailable in the mainstream setting: the agents consider a whole family $\mathcal{S}$ of stochastic processes, \emph{defined on the same fixed stochastic basis}, as the class of possible scenarios for the prices, all of which are seen as equally plausible. %
Put differently, while in our context investors know the real-world likelihood of outcomes (in particular, what the null and sure events are), they are uncertain about the dynamics followed by the prices, so the misspecification of the true model is represented by a collection of processes to consider. %
In short, the present setting deviates from the usual one in that, instead of fixing a process $S$ on a measurable space $\left(\Omega,\mathscr{F}\right)$ (in most cases $S$ consists of the coordinate mappings on a convenient path space) and considering a family $\mathscr{P}$ of probabilities on $\left(\Omega,\mathscr{F}\right)$, we fix the stochastic basis (i.e.\ the available information together with the true chances of events occurring) and vary the process $S\in\mathscr{S}$. %
Parametric classes of models, which assume a particular specification for the evolution of prices with some common driving factor structure and a set of (unknown or unspecified) parameters determining the price dynamics, typically fall into our setting. %

In the standard setting, robust utilities give rise to preference relations between all (suitably integrable) random variables (see e.g.\ the axiomatic treatment in \citet{gilboaschmeidler1989}), in particular preferences are defined for every possible portfolio terminal wealth. In our setting, on the other hand, preferences are actually defined on agents' \emph{decisions} (specifically, the admissible controls/investment policies/portfolio strategies), where the utility of a decision (portfolio strategy) is the infimum of its expected utility over the family of possible models. %
Apparently there is no way to extend these preferences to \emph{all} random variables. %
There is no obvious relationship between these two approaches. %
Ours emphasises that there is a class of models (e.g.\ corresponding to a parametrisation), the investor is uncertain about which is the right one and prefers strategies with higher worst-case utilities. %
In the standard approach uncertainty is already present at the choice of the probability space, as usual in theoretical statistics, and any two random outcomes can be compared, regardless of being a portfolio terminal value or not. %

An important aspect to notice is that this alternative uncertainty modelling approach does not exclude the laws of these processes being e.g.\ mutually singular. %
One obvious comparative advantage at the core of our work is that we are still able to use results of classical probability theory; namely, we are able to avoid the delicate measurability issues arising in the non-dominated quasi-sure framework (we point out that we do not impose any particular structure on our family of models $\mathscr{S}$), and apply Koml\'{o}s-type arguments to construct candidates for optimal strategies. %
In this way, we can cover certain cases which the standard setting cannot. %
There are also model specifications that elude our present approach, but are conveniently handled with the existing tools. Hence, our contribution complements, without subsuming, the prevailing approach of e.g.\ \citet{nutz2016,bartl2017,blanchard-carassus2017,neufeld-sikic2018}; see \Cref{sec:Example} for more on this. %

The aim of this paper is to investigate conditions under which the robust utility maximisation problem from terminal wealth for a risk-averse investor in a discrete-time financial market admits a solution. %
The agents adopt a worst-case approach in evaluating a given payoff by first minimising the welfare performance of each portfolio over all the stock prices that can materialise, and only then selecting the investment strategies that offer the best of such worst-scenario utilities. %
In the aforementioned papers in the non-dominated quasi-sure setting, a main mathematical tool for proving the existence of optimal strategies is dynamic programming, which allows to reduce the multi-period robust portfolio choice problem to a sequence of one-period decision problems, but leads to difficult measurability and analyticity issues (for instance, at each time step it is necessary to consider a suitable family of possible transition probability measures, and subsequently take their concatenation; this is referred to as ``time-consistency'' in the terminology of \citet{bck2019}). In the present work, we instead use an approach that deals with the primal problem directly, by constructing optimal portfolios from optimising sequences (i.e., whose worst-case expected utility becomes arbitrarily close to the supremum over all strategies). %
We need neither the time-consistency assumption (present in all related papers except \citet{bck2019}) nor extra set-theoretical hypotheses (present in \citet{bck2019}).

Our results are obtained under the assumption that there are no arbitrage opportunities for at least \emph{some} price evolution in the collection of all possible models and that model is non-degenerate in a suitable sense. %
We compare this condition to the conventional robust no-arbitrage condition in \Cref{obs:RNA} below. %
This condition enables us to work with an equivalent martingale measure (whose existence would otherwise be unknown at this time); it is also intimately connected to other existing notions of absence of arbitrage, such as the ``weak no-arbitrage'' first proposed by \citet{davishobson2007} (see \Cref{as:NA} and the discussion in \Cref{obs:RNA} below). %

One main contribution of this work is that utility functions on the positive axis and on the whole real line are both treated. In addition, we consider bounded above functions as well as functions that can grow without bound; in the latter case, we must replace boundedness of the utility with certain integrability conditions on quantities related to the price process. %
Random endowments or liabilities could be admitted at little cost, but this is not pursued in the present work. %

The structure of the paper is as follows: \Cref{sec:Model} presents the model and the mathematical formulation of the utility maximisation problem taking into account model uncertainty; \Cref{sec:Main} states and discusses our main results on the existence of robust optimal strategies; \Cref{sec:Example} contains some illustrative examples highlighting important differences, advantages, and drawbacks of the present framework in relation to the quasi-sure framework; \Cref{sec:Conc} outlines future directions of research and concludes. For fluidity, all proofs are collected in the \hyperref[app:Main]{Appendix}. %


\section{Model}\label[sec]{sec:Model}


Consider a financial market consisting of $d\in\mathbb{N}$ traded risky assets plus an additional risk-free asset with constant unit price, and fix a time horizon $T\in\mathbb{N}$. %
Let $(\Omega,\mathscr{F},\mathbb{F}=\left\{\mathscr{F}_{t}\right\}_{t\in\left\{0,1,\ldots,T\right\}},\mathbb{P})$ be a discrete-time stochastic basis, and assume that the $\sigma$-algebras occurring in this paper contain all $\mathbb{P}$-null sets. %
For every $n\in\mathbb{N}$ and $\mathscr{G}\subseteq\mathscr{F}$, denote by $\Lspace[n]{0}{}{\mathscr{G}}$ the space of $\mathscr{G}$-measurable, $\mathbb{R}^{n}$-valued random variables (as usual, we identify random variables coinciding outside a $\mathbb{P}$-null set), endowed with the metrisable topology of convergence in probability. %

Let $\mathscr{S}$ be a non-empty family of adapted, $\mathbb{R}^{d}$-valued processes on this stochastic basis, and define the topological product space $\mathscr{L}\coloneqq\otimes_{t=1}^{T}L^{0}(\mathscr{F}_{t-1};\mathbb{R}^{d})$. %
When the (discounted) prices of the risky assets evolve according to the process $S\in\mathscr{S}$, the (discounted) wealth at time $t\in\left\{0,1,\ldots,T\right\}$ of a self-financing portfolio $\phi\in\mathscr{L}$ starting from initial capital $\iw\in\mathbb{R}$ is given by
\begin{equation*}
	W_{t}^{S}\!\left(\iw,\phi\right)=\iw+\sum_{s=1}^{t}\innerp{\phi_{s}}{\Delta S_{s}}.
\end{equation*}
Here, $\Delta S_{t}\coloneqq S_{t}-S_{t-1}$ is the price change at trading period $t\in\left\{1,\ldots,T\right\}$, while $\innerp{\cdot}{\cdot}$ denotes the Euclidean inner product in $\mathbb{R}^{d}$ with the corresponding norm $\norm{\cdot}$. %
A strategy is admissible if it is feasible for every particular possible price; more precisely, the set of admissible trading strategies with initial capital $\iw$ is %
	$\adm{\iw}\coloneqq\bigcap_{S\in\mathscr{S}}\mathscr{L}\!\left(\iw,S\right)%
	$%
\ for some $\mathscr{L}\!\left(\iw,S\right)\subseteq\mathscr{L}$ to be specified later, depending on the domain of the utility function.

Next, for every $S\in\mathscr{S}$ and $t\in\left\{1,\ldots,T\right\}$, let $\rcd[t-1]{\Delta S_{t}}{F}\!\left(\cdot,\cdot\right)$ be a regular conditional distribution of $\Delta S_{t}$ with respect to $\mathscr{F}_{t-1}$. By redefining on a $\mathbb{P}$-null set, $\rcd[t-1]{\Delta S_{t}}{F}\!\left(\cdot,\omega\right)$ is a probability measure on $(\mathbb{R}^{d},\mathscr{B}(\mathbb{R}^{d}))$ for all $\omega\in\Omega$, and we denote by $D_{t}^{S}\!\left(\omega\right)$ the smallest affine subset of $\mathbb{R}^{d}$ containing the support of $\rcd[t-1]{\Delta S_{t}}{F}\!\left(\cdot,\omega\right)$.

Throughout the paper, assume that the collection of stock prices contains at least one process for which there is no chance of making a riskless profit out of nothing. In addition, the redundant assets in the trading mechanism for such price model must also be redundant for all the other price processes that can materialise. These requirements are formalised as follows.
\begin{assumption}\label[as]{as:NA}
	There exists $S^{*}\in\mathscr{S}$ satisfying both
	\begin{equation*}\label{eq:NA}\labeltarget{eq:NAS}
		\text{for all }\phi\in\mathscr{L}\!\left(\iw,S^{\ast}\right),\ \text{if }W^{S^{\ast}}_{T}\!\left(0,\phi\right)\geq0\ \mathbb{P}\text{-almost surely (a.s.)},\ \text{then }W^{S^{\ast}}_{T}\!\left(0,\phi\right)=0\ \mathbb{P}\text{-a.s.}%
			,%
		\tag{NA($S^{\ast}$)}
	\end{equation*}
	and for each $S\in\mathcal{S}$%
	\begin{equation}\label{eq:nonred}
		D_{t}^{S}\subseteq D_{t}^{S^{\ast}}\text{almost surely, for all }t\in\left\{1,\ldots,T\right\}.
	\end{equation}
	The set of such $S^{*}$ will be denoted by $\NAset$.
\end{assumption}

The following alternative characterisation of the arbitrage-free and ``non-redundant'' price processes is provided by \citet[Proposition~3.3]{rs05} (see also \citet[Proposition~2.1]{carassusrasonyi2009}).
\begin{samepage}
\begin{proposition}\label[prop]{prop:NA}
	Let $S\in\mathscr{S}$. %
	The two statements below are equivalent.\nopagebreak
	\begin{enumerate}[label=\emph{(\roman*)}]
		\item
		\hyperlink{eq:NAS}{NA($S$)} holds.
		
		\item
		For every $t\in\left\{1,\ldots,T\right\}$, there exist $\mathscr{F}_{t-1}$-measurable random variables $\beta_{t}^{S}>0%
		$ $\mathbb{P}$-a.s.\ and $\kappa_{t}^{S}>0$ $\mathbb{P}$-a.s.\ such that
		\begin{equation}\label{eq:altNA}
			\pushQED{\qed}%
			\essinf_{\xi\in\Xi_{t}^{S}}%
			\mathbb{P}\!\left\{\left.\innerp{\xi}{\Delta S_{t}}\leq-\beta_{t}^{S}\norm{\xi}\right\vert\mathscr{F}_{t-1}\right\}\geq\kappa_{t}^{S}\ \mathbb{P}\text{-a.s.\ on %
			} \left\{D_{t}^{S}\neq\left\{\vect{0}\right\}\right\}%
			, %
		\end{equation}
		where $\Xi_{t}^{S}\coloneqq\left\{\xi\in\Lspace[d]{0}{}{\mathscr{F}_{t-1}}:~\xi\!\left(\omega\right)\in D_{t}^{S}\!\left(\omega\right)\text{ for }\mathbb{P}\text{-a.e.\ }\omega\in\Omega\right\}$.
	\end{enumerate}
\end{proposition}
\end{samepage}

Furthermore, we make the assumption that the investors' risk preferences are continuous and non-satiable, with declining marginal utility. %
These are fairly weak conditions on the utility function: we require neither strict concavity nor smoothness (in particular, no Inada-type conditions need to be satisfied).
\begin{assumption}\label[as]{as:U}
	The \emph{utility} function $U:~\mathbb{R}\rightarrow\mathbb{R}\cup\left\{-\infty\right\}$ is non-decreasing, concave and upper semicontinuous.
\end{assumption}

We define
\begin{equation}\label{eq:domU}
	\dom\!\left(U\right)\coloneqq\left\{x\in\mathbb{R}:~U\!\left(x\right)>-\infty\right\},
\end{equation}
which is an interval of the form $\left(b,+\infty\right)$ or $\left[\left.b,+\infty\right)\right.$ for some $b\in\mathbb{R}\cup\{\pm\infty\}$. %
Upper semicontinuity means that, if $b\in\mathbb{R}$, then $\lim_{x\rightarrow b^{+}}U\!\left(x\right)=U\!\left(b\right)$ (where the latter may be finite or $-\infty$).

In the presence of uncertainty, the investors' objective is to choose an admissible portfolio that maximises their expected utility from terminal wealth in the worst possible concretisation of the asset prices, where the same plausibility is attached to every price process in $\mathscr{S}$. This model uncertainty version of the optimal portfolio problem is defined precisely below. %
\begin{definition}[Robust optimal portfolio]\label[def]{def:rEU}
	A portfolio $\opt\in\adm{\iw}$ is \emph{optimal} under model uncertainty for an investor with initial wealth $\iw$ if
	\begin{equation}\label{eq:rEU}
		u\!\left(\iw\right)\coloneqq\sup_{\varphi\in\adm{\iw}}\inf_{S\in\mathscr{S}}\mathbb{E}_{\mathbb{P}}\!\left[U\!\left(W_{T}^{S}\!\left(\iw,\varphi\right)\right)\right]=\inf_{S\in\mathscr{S}}\mathbb{E}_{\mathbb{P}}\!\left[U\!\left(W_{T}^{S}\!\left(\iw,\opt\right)\right)\right].
	\end{equation}
	Here, we adopt the usual convention that, if a random variable is such that its positive and negative parts both have infinite expectation, then its expectation is defined to be $-\infty$ (i.e., $+\infty-\infty\coloneqq-\infty$).
\end{definition}

\begin{remark}\label[obs]{obs:u>-infty}
	It is trivial that %
		$
		\inf_{S\in\mathscr{S}}\mathbb{E}_{\mathbb{P}}\!\left[U\!\left(W_{T}^{S}\!\left(\iw,\safe\right)\right)\right]=U\!\left(\iw\right)$%
	\ for all $\iw\in\dom\!\left(U\right)$%
	, where $\safe\equiv\left(\vect{0},\ldots,\vect{0}\right)$ denotes the safe portfolio allocating all wealth to the safe asset. %
	Therefore, $u\!\left(\iw\right)>-\infty$ provided that the safe portfolio is admissible. %
\end{remark}

The next section demonstrates the existence of an optimiser for \eqref{eq:rEU} in two settings: initially, investors are subject to the usual no-bankruptcy requirement; subsequently, we relax this restriction by treating the case of finite utility on the whole real line.

\begin{remark}\label[obs]{obs:RNA}
	\mbox{}
	\begin{enumerate}[label=\emph{(\roman*)}]
	
	\item
	We establish our results under \Cref{as:NA}, which implies that there is at least \emph{one} model $S^{\ast}\in\mathscr{S}$ that is arbitrage-free in the classical sense; in other words, \hyperlink{eq:NAS}{NA($S^{\ast}$)} corresponds to the classical absence of arbitrage opportunities for the price process $S^{\ast}$. %
	Incidentally, this condition is the counterpart of the conventional concept of ``weak no-arbitrage'' of \citet{davishobson2007} (itself a particular case of the ``no $\mathscr{R}$-arbitrage'' property formulated in \citet[Definition~5.1]{bfm2016}). %
	
	\item
	Another pertinent comparison is with the standard ``robust no-arbitrage condition'' introduced by \citet[Definition~1.1]{bouchardnutz2015}, which in the present framework reads as:
	\begin{equation*}\label{eq:RNA}
		\text{For all }S\in\mathscr{S},\ W_{T}^{S}\!\left(0,\phi\right)\geq0\ \mathbb{P}\text{-a.s.}\Rightarrow \text{For all }S\in\mathscr{S},\ W_{T}^{S}\!\left(0,\phi\right)=0\ \mathbb{P}\text{-a.s.}.\tag{
		{NA($\mathscr{S}$)}%
		}
	\end{equation*}
	
	In the multiple priors setting, \citet[Theorem 3.8]{bc19} show that (the multiple-prior equivalent of) \ref{eq:RNA} implies weak no-arbitrage, but the converse direction fails; refer to \citet[statement (5) of Lemma~3.7]{bc19}. %
	In addition, Theorem~3.30 of \citet{bc19} states that (the multiple-prior equivalent of) \Cref{as:NA} is actually equivalent to the robust no-arbitrage condition of \citet{bouchardnutz2015}, hence it seems to be a fairly natural requirement. %
	Note, however, that the multiple priors framework is different from the current one, therefore it is not possible to directly transfer results from either one to the other. %
	
	Returning to our framework, \Cref{as:NA} is stronger than condition~\ref{eq:RNA}. %
	To see this, let $\phi\in\adm{\iw}$ such that, for all $S\in\mathscr{S}$, it holds $\mathbb{P}$-a.s.\ that $W_{T}^{S}\!\left(0,\phi\right)\geq0$. Then, for all $t\in\left\{1,\ldots,T\right\}$, we get $W_{t}^{S^{\ast}}\!\left(0,\phi\right)=0$ $\mathbb{P}$-a.s., which in turn implies that $\phi_{t}\!\left(\omega\right)$ belongs to the orthogonal complement of $D_{t}^{S^{\ast}}\!\left(\omega\right)$ for $\mathbb{P}$-a.e.\ $\omega\in\Omega$. Hence, $\phi_{t}$ is also orthogonal to $D_{t}^{S}\subseteq D_{t}^{S^{\ast}}$ for all $S\in\mathcal{S}$, and we conclude $W_{T}^{S}\!\left(0,\phi\right)=0$ $\mathbb{P}$-a.s. %
	Unlike in the multiple priors case,	however, in the present framework \ref{eq:RNA} does not imply in general that \Cref{as:NA} holds or even \hyperlink{eq:NAS}{NA($S^{\ast}$)} for some $S^{\ast}$. Indeed, consider a one-period model with one risky asset (i.e.\ take $T=1$ and $d=1$) where $\mathscr{S}$ consists of two processes: $\bar{S}$ such that $\mathbb{P}\!\left\{\Delta\bar{S}_{1}=0\right\}=\mathbb{P}\!\left\{\Delta\bar{S}_{1}=1\right\}=1/2$, and $\tilde{S}$ such that $\mathbb{P}\!\left\{\Delta\tilde{S}_{1}=0\right\}=\mathbb{P}\!\left\{\Delta\tilde{S}_{1}=-1\right\}=1/2$; then although \ref{eq:RNA} holds, neither \hyperlink{eq:NAS}{NA($\bar{S}$)} nor \hyperlink{eq:NAS}{NA($\tilde{S}$)} do (note also that $D^{\bar{S}}_{1}=D^{\tilde{S}}_{1}=\mathbb{R}$)%
	.\footnote{We thank the referee for suggesting this counter-example.} %
	
	The reason why we work under the ``weak no-arbitrage'' assumption rather than the ``robust no-arbitrage'' condition is that the former provides us with an equivalent martingale measure, by the Dalang-Morton-Willinger theorem~\citep{dalang1990}; obtaining a version of the Fundamental Theorem of Asset Pricing associated with \ref{eq:RNA} is a project worth further pursuing in the current setting (as already done in the quasi-sure framework; see \citet{bouchardnutz2015,bc19}). %
	
	\item
	Compare also to \citet{blanchard-carassus2017}, where unbounded from above utilities defined on the positive half-real line are treated under a (strengthened) no-arbitrage condition for \emph{every} model $S\in\mathscr{S}$ (see Definition~2.4 therein). %
	We point out that we obtain our existence \Cref{th:UrealU} below, for unbounded utilities over the whole real line, under a similar (strong) absence of arbitrage assumption for all possible price processes. This trivially implies the robust no-arbitrage condition~\ref{eq:RNA}, 
	as also observed by \citet[Lemma~3.7(4)]{bc19} in the multi-priors setting; additionally, see Theorem~3.8 in the same paper, establishing the equivalence between robust no-arbitrage and the existence of some subfamily of priors (having the same ``relevant'' events as the original one) which are all arbitrage-free. %
	It is still unclear what the suitable notion of no-arbitrage should be in addressing the unbounded robust optimal portfolio problem. %
	Recall that, in utility maximisation ``under certainty'', absence of arbitrage is a necessary condition for the existence of an optimiser, as shown e.g.\ by \citet[Proposition~3.1]{rs05}. %
	\end{enumerate}
\end{remark}


\section{Main Results}\label[sec]{sec:Main}

The proofs of the existence theorems in this section all share a common structure: First, we verify the finiteness of the value function for the optimisation problem~\eqref{eq:rEU}, so as to ensure that investors cannot attain near bliss from the strategies available in the market%
. We can then take a maximising sequence of portfolios, and obtain from it a candidate for an optimiser via a compactness argument. Lastly, we use the upper semicontinuity of the worst-case expected utility with respect to the strategies to check that the portfolio found in the preceding step is indeed optimal. %

Under the upper boundedness assumption on risk preferences, both well-posedness and up\-per semi\-con\-ti\-nu\-i\-ty are straightforward%
---the only point remaining to show being that a suitable collection of strategies satisfies some form of compactness. %
Once we drop the requirement that the utility should be bounded from above, however, we start running into some integrability problems (namely, we no longer have an obvious integrable upper bound for the positive part of terminal utility, guaranteeing the convergence of certain integrals), which we handle by introducing additional hypotheses. %

When the non-bankruptcy constraint is in force, the classical absence of arbitrage combined with trading ``non-redundancy'' in \emph{one} of the realisable models is enough to establish bounds on the admissible strategies (given by the auxiliary, almost surely finite random variables $G^{S}_{t}$ found in \Cref{lem:key} below). %
For the reasons just set forth, with unbounded utilities we need these bounds to be integrable as well, which is achieved through a strengthening of the no-arbitrage condition, in the sense of~\eqref{eq:intW}. %

Turning to a setting where wealth is allowed to take on positive as well as negative values with positive probability, we again see that boundedness of preferences together with risk aversion suffices to have compactness; the arguments are more lengthy and involved than those of its positive real line counterpart, since the bounds of \Cref{lem:key} are no longer available. %
Finally, for the particularly delicate case of an unbounded utility defined on the whole real line, the proof becomes even more intricate; as it turns out, absence of arbitrage in just one model is not enough anymore, and we must actually assume that \emph{every} possible model admits no arbitrage opportunities. On top of it, even though we allow the utility to grow arbitrarily large, we must require that it does so
in a strictly sublinear way in the sense of \eqref{maruszja} below. These conditions ensure the upper semicontinuity properties that are needed for the proof.

Summing up, if we consider utilities that are bounded above, then we are able to show that a robust optimal portfolio exists without having to impose any conditions other than \Cref{as:NA}; by contrast, the robust portfolio choice with an unbounded utility involves increased difficulties that call for considerably more stringent assumptions, not only on arbitrage but also on price dynamics and utility growth. These are not surprising, see the related discussions in \citet{nutz2016}, and \citet{blanchard-carassus2017}. %

\subsection{Utility on the positive real line}\label[subsec]{subsec:positiverl}
We start by focusing on the case where debt is disallowed under every scenario for the evolution of stock prices, so the utility function has effective domain of definition equal to the positive half-line. %
\begin{assumption}\label[as]{as:positiverl}
	For all $S\in\mathscr{S}$,
	\begin{equation}\label{eq:LS}
		\mathscr{L}\!\left(\iw,S\right)\coloneqq\left\{\phi\in\mathscr{L}:~%
		\forall\,%
		t\in\left\{0,1,\ldots,T\right\},\ W_{t}^{S}\!\left(\iw,\phi\right)\geq0\ \mathbb{P}%
		\text{-a.s.}%
		\right\}.
	\end{equation}
	Moreover, $\inter\!\left(\dom\!\left(U\right)\right)=\left(0,+\infty\right)$.
\end{assumption}

\begin{remark}
	Due to monotonicity, $U\!\left(\cdot\right)$ extends in a natural way to $\left[\left.0,+\infty\right)\right.$ by setting %
		$U\!\left(0\right)\coloneqq\lim_{x\rightarrow0^{+}}U\!\left(x\right)$
	, which is possibly $-\infty$. %
	Suppose further, without loss of generality, that $U\!\left(+\infty\right)\coloneqq\lim_{x\rightarrow+\infty}U\!\left(x\right)>0$. %
	Finally note that, in addition to being non-empty and convex, $\adm{\iw}$ is closed with respect to the subspace topology $\tau$ of the product topology on $\mathscr{L}$.
\end{remark}

Both existence theorems of the current subsection rely on the following key lemma, whose proof proceeds along the lines of that of \citet[Lemma~2.1]{rsch06}. This result comes into play in the construction of an optimal portfolio as it implies the boundedness of the admissible set $\adm{\iw}$, therefore making it possible to employ a compactness argument to extract a candidate solution to~\eqref{eq:rEU}.
\begin{samepage}
\begin{lemma}\label[lem]{lem:key}
	Let $\iw>0$, and let \Cref{as:NA} hold. Moreover, let \Cref{as:positiverl%
	} hold. %
	Then, for every $S\in\NAset$ and $t\in\left\{1,\ldots,T\right\}$, there exists an $\mathscr{F}_{t-1}$-measurable random variable $G^{S}_{t}>0$ $\mathbb{P}$-a.s.\ such that
	\begin{equation}\label{eq:etoile}
		\esssup_{\phi\in\mathscr{L}\!\left(\iw,S\right)}\norm{\hat{\phi}_{t}^{S}}\leq G^{S}_{t}<+\infty\ \mathbb{P}\text{-a.s.}%
		, %
	\end{equation}
	where $\hat{\phi}_{t}^{S}\!\left(\omega\right)$ denotes the orthogonal projection of $\phi_{t}\!\left(\omega\right)$ on $D_{t}^{S}\!\left(\omega\right)$ for all $\omega\in\Omega$.
\end{lemma}

\begin{proof}
	See~\Cref{app:Main}
	.
\end{proof}
\end{samepage}

\begin{remark}\label[obs]{obs:proj}
	For all $\phi\in\mathscr{L}\!\left(\iw,S\right)$ and $t\in\left\{1,\ldots,T\right\}$, we have $\hat{\phi}_{t}^{S}\in\Lspace[d]{0}{}{\mathscr{F}_{t-1}}$ as well as $\innerp{\hat{\phi}_{t}^{S}}{\Delta S_{t}}=\innerp{\phi_{t}}{\Delta S_{t}}$ $\mathbb{P}$-a.s.\ (refer to \citet[Remark~3.4]{carassusrasonyi2012}), so replacing a portfolio with its orthogonal projection does not alter its value. More importantly, when Assumption \ref{as:NA} holds	and $S^{\ast}\in\mathcal{S}^{\ast}$	then, for all $S\in\mathcal{S}$, we have $\innerp{\hat{\phi}_{t}^{S^{\ast}}}{\Delta S_{t}}=\innerp{\phi_{t}}{\Delta S_{t}}$	almost surely since $D^{S}\subseteq D^{S^{\ast}}$. %
\end{remark}

Our first main result states that, whenever the investors' utility is bounded from above, not only is the robust utility maximisation problem~\eqref{eq:rEU} trivially well-posed (i.e., $u\!\left(\iw\right)<+\infty$ for every initial capital $\iw$), but an optimal strategy always exists.
\begin{theorem}\label[th]{th:Ubounded}
	Let $\iw>0$, and let \Cref{as:NA,as:U,as:positiverl%
	} hold. If %
		$U\!\left(+\infty\right)%
		<+\infty$, %
	then there exists $\opt\in\adm{\iw}$ such that 
	\begin{equation*}
		u\!\left(\iw\right)=\inf_{S\in\mathscr{S}}\mathbb{E}_{\mathbb{P}}\!\left[U\!\left(W_{T}^{S}\!\left(\iw,\opt\right)\right)\right]%
		<+\infty%
		. %
	\end{equation*}
\end{theorem}

\begin{proof}
	See~\Cref{app:Main}
	.
\end{proof}

Alternatively, suppose that investors can derive an arbitrarily high level of satisfaction as wealth grows arbitrarily large. %
In order to still be able to control from above the robust expected utility of terminal wealth, and thus obtain a solution for~\eqref{eq:rEU}, we make the additional assumption~\eqref{eq:intW} concerning the dynamics of \emph{all} possible prices combined with a stronger version of absence of arbitrage in at least \emph{one} of the models. %
\begin{theorem}\label[th]{th:intW}
	Let $\iw>0$, and let \Cref{as:NA,as:U,as:positiverl%
	} be valid. %
	Also, define %
	\begin{equation}\label{eq:W}
		\mathscr{W}\coloneqq\left\{X\in\Lspace[]{0}{}{\mathscr{F}_{T}}:~\forall\,p>0,\ \mathbb{E}_{\mathbb{P}}\!\left[\left\vert X\right\vert^{p}\right]<+\infty\right\}.
	\end{equation}
	Assume further that 
	\begin{equation}\label{eq:intW}
		\left\{\norm{\Delta S_{t}}:~S\in\mathscr{S}\right\}\subseteq\mathscr{W}%
		\text{ and }%
		\frac{1}{\beta^{S^{\ast}}_{t}}\in\mathscr{W}\ \text{for all }t\in\left\{1,\ldots,T\right\}
	\end{equation}
	holds for some $S^{\ast}\in\NAset$, where each $\beta^{S^{\ast}}_{t}$ is the random variable given by \Cref{prop:NA}. %
	Then there exists $\opt\in\adm{\iw}$ such that 
	\begin{equation*}
		u\!\left(\iw\right)=\inf_{S\in\mathscr{S}}\mathbb{E}_{\mathbb{P}}\!\left[U\!\left(W_{T}^{S}\!\left(\iw,\opt\right)\right)\right]%
		<+\infty%
		. %
	\end{equation*}
\end{theorem}

\begin{proof}
	See~\Cref{app:Main}
	.
\end{proof}

\begin{remark}
	We may wonder what happens if we relax the bankruptcy prohibition at all intermediate times by imposing that the admissible strategies for each price process $S\in\mathscr{S}$ yield non-negative wealth only at the terminal time.
	\ In other words, instead of $\adm{\iw}$, we can consider the larger set
	\begin{equation*}
		\admE{\iw}\coloneqq\bigcap_{S\in\mathscr{S}}\tilde{\mathscr{L}}\!\left(\iw,S\right),
	\end{equation*}
	where
	\begin{equation}\label{eq:LE}
		\tilde{\mathscr{L}}\!\left(\iw,S\right)\coloneqq\left\{\phi\in\mathscr{L}:~W_{T}^{S}\!\left(\iw,\phi\right)\geq0\ \mathbb{P}%
		\text{-a.s.}%
		\right\}.
	\end{equation}
	Clearly, for all $\iw>0$,
	\begin{equation*}
		\tilde{u}\!\left(\iw\right)\coloneqq\sup_{\varphi\in\admE{\iw}}\inf_{S\in\mathscr{S}}\mathbb{E}_{\mathbb{P}}\!\left[U\!\left(W_{T}^{S}\!\left(\iw,\varphi\right)\right)\right]\geq u\!\left(\iw\right),
	\end{equation*}
	so the question arises of whether the above inequality can be strict. %
	
	It is not difficult to see that, under \Cref{as:NA}, %
	$\tilde{\mathscr{L}}\!\left(\iw,S\right)=\mathscr{L}\!\left(\iw,S\right)$ for all $S\in\NAset$%
	. Indeed, if we let $\phi\in\mathscr{L}$ with $W_{T}^{S}\!\left(\iw,\phi\right)\geq0$ $\mathbb{P}$-a.s., then we know by the Dalang-Morton-Willinger theorem~\citep{dalang1990} combined with Theorem~2 of \citet{jacod98} that the wealth process $\left\{W^{S^{\ast}}_{t}\!\left(\iw,\phi\right)\right\}_{t\in\left\{0,1,\ldots,T\right\}}$ is a true martingale under some equivalent martingale measure. The desired conclusion follows from the observation that a martingale with non-negative terminal value is non-negative. %

	Additionally, we can easily check that, if we replace $\adm{\iw}$ with $\admE{\iw}$ in \Cref{th:Ubounded,th:intW}, the results remain valid with the same proofs (the key fact being that the two sets of feasible portfolios actually coincide for all arbitrage-free price models, as shown above). However, the value of the robust utility maximization problem over these two domains may well differ; below we provide an example where this happens. %
	This is in sharp contrast with the multiple-priors setting, where such a relaxation of the admissible set does not affect the optimisation problem, since under robust no-arbitrage a quasi-surely non-negative terminal wealth implies the quasi-sure non-negativity of wealth at all intermediate times (refer e.g.\ to \citet[Lemma~4.3]{blanchard-carassus2017}).\footnote{We thank the referee for posing this interesting question, as well as for pointing out another fundamental difference between the two frameworks.}

	Let $T\coloneqq2$, and let $X$ be a standard Gaussian. Consider also the random variables $\varepsilon_{i}$, $i\in\left\{1,2\right\}$ such that
	\begin{align*}
		&\mathbb{P}\!\left\{\varepsilon_{1}=-1/2\right\}=\mathbb{P}\!\left\{\varepsilon_{1}=4\right\}=1/2,\\
		&\mathbb{P}\!\left\{\varepsilon_{2}=-1/2\right\}=\mathbb{P}\!\left\{\varepsilon_{2}=1/2\right\}=1/2,
	\end{align*}
	and $X,\varepsilon_{1},\varepsilon_{2}$ are independent. Moreover, the filtration $\mathbb{F}$ is defined by setting $\mathscr{F}_{0}$ to be the family of $\mathbb{P}$-null sets, $\mathscr{F}_{1}\coloneqq\mathscr{F}_{0}\vee\sigma\!\left(X,\varepsilon_{1}\right)$, and $\mathscr{F}_{2}\coloneqq\mathscr{F}_{1}\vee \sigma(\varepsilon_{2})$.\footnote{Here, $\mathscr{G}\vee\mathscr{H}$ denotes the $\sigma$-algebra generated by the union of the two $\sigma$-algebras $\mathscr{G}$ and $\mathscr{H}$.}%

	We consider one risky asset, i.e.\ $d=1$. %
	Our family of models has $3$ elements, $\mathscr{S}=\left\{S^{\ast},\tilde{S},\bar{S}\right\}$, defined as follows,
	\begin{alignat*}{3}
		&\Delta S^{\ast}_{1}= \varepsilon_{1},\qquad &&\Delta S^{\ast}_{2}= \varepsilon_{2},\\
		&\Delta \tilde{S}_{1}= X,\qquad &&\Delta \tilde{S}_{2}= 3-X,\\
		&\Delta \bar{S}_{1}= 3,\qquad &&\Delta \bar{S}_{2}= 0.
	\end{alignat*}
	Clearly, $D_{1}^{S^{\ast}}=D_{2}^{S^{\ast}}=\mathbb{R}$, hence \Cref{as:NA} is testified by $S^{\ast}\in\NAset$. %

	Take the bounded utility $U\!\left(x\right)\coloneqq\min\{\sqrt{x},2\}$, for all $x\in\left(0,+\infty\right)$, and set the initial capital to $\iw=1$. Each portfolio strategy $\phi=(\phi_{1},\phi_{2})$ consists of a deterministic number $\phi_{1}\in\mathbb{R}$, and an $\mathscr{F}_{1}$-measurable real-valued random variable $\phi_{2}$. %

	Notice that $\phi\in\adm{1}$ implies $\phi_{1}=0$, since $X$ is unbounded both from above and from below. %
	Hence, by the construction of $\bar{S}$,
	\begin{equation*}
		u\!\left(1\right)=%
		\sup_{\phi\in\adm{1}}\inf_{S\in\mathscr{S}}\mathbb{E}_{\mathbb{P}}\!\left[U\!\left(W_{T}^{S}\!\left(1,\phi\right)\right)\right]%
		\leq\sup_{\phi\in\adm{1}}\mathbb{E}_{\mathbb{P}}\!\left[U\!\left(W_{T}^{\bar{S}}\!\left(1,\phi\right)\right)\right]%
		=U\!\left(1\right)=1.
	\end{equation*}

	On the other hand, choosing the strategy $\tilde{\phi}$ with $\tilde{\phi}_{1}=\tilde{\phi}_{2}=1$ gives $W_{T}^{S}\!\left(1,\tilde{\phi}\right)\geq 0$ for all $S\in\mathscr{S}$, thus $\tilde{\phi}\in\admE{1}\setminus\adm{1}$. Moreover, we can directly check that, for all $S\in\mathscr{S}$,
	\begin{equation*}
		\mathbb{E}_{\mathbb{P}}\!\left[U\!\left(W_{T}^{S}\!\left(1,\tilde{\phi}\right)\right)\right]\geq\frac{5}{4}%
		, %
	\end{equation*}
	which in turn results in
	\begin{equation*}
		\tilde{u}\!\left(1\right)=%
		\sup_{\phi\in\admE{1}}\inf_{S\in\mathscr{S}}\mathbb{E}_{\mathbb{P}}\!\left[U\!\left(W_{T}^{S}\!\left(1,\phi\right)\right)\right]>1.
	\end{equation*}
\end{remark}

\subsection{Utility on the whole real line}\label[subsec]{subsec:wholerl}

Consider now the possibility that wealth can become negative, meaning that the utility function takes on finite values everywhere on the real line. In other words, no constraints are imposed on portfolio choice, hence any self-financing investment strategy is admissible.
\begin{assumption}\label[as]{as:wholerl}
	For all $S\in\mathscr{S}$, %
		$\mathscr{L}\!\left(\iw,S\right)\coloneqq\mathscr{L}$%
		. %
	Moreover, $\dom\!\left(U\right)=\mathbb{R}$.
\end{assumption}%

The next result extends our \Cref{th:Ubounded} to the whole real line and shows that there is also an optimal portfolio in this setting.
\begin{theorem}\label[th]{th:Ureal}
	Let $\iw\in\mathbb{R}$, and let \Cref{as:NA,as:U,as:wholerl%
	} hold. %
	If %
		$U\!\left(+\infty\right)%
		<+\infty$, %
	then there exists $\opt\in\mathscr{L}$ such that 
	\begin{equation*}
		u\!\left(\iw\right)%
		=\inf_{S\in\mathscr{S}}\mathbb{E}_{\mathbb{P}}\!\left[U\!\left(W_{T}^{S}\!\left(\iw,\opt\right)\right)\right]%
		<+\infty%
		. %
	\end{equation*}
\end{theorem}

\begin{proof}
	See~\Cref{app:Main}
	.
\end{proof}

In our last theorem, we remove the hypothesis of $U\!\left(\cdot\right)$ being bounded from above. This comes at the expense of assuming that \emph{each} $S\in\mathscr{S}$ should be arbitrage-free, in a strong sense; we further point out that \eqref{eq:lune} bears a close resemblance to the conditions of \Cref{th:intW} above as well as those of \citet[Theorem~3.6]{blanchard-carassus2017}. %
Likewise, dealing with the integrability issues that arise from $U\!\left(+\infty\right)=+\infty$ requires a supplementary growth condition on the utility for large positive values of wealth---namely, that the function displays strictly sub-linear growth. %
Nonetheless, \eqref{maruszja} is still quite mild; in fact, it is even slightly weaker than the reasonable asymptotic elasticity property at $+\infty$ (see \citet{kramkov99}). %
As far as we know, this is the first result in the literature that treats, in a robust setting, an unbounded utility function defined on the whole real line.

\begin{remark}\label[obs]{obs:AdmR}
	A simple, yet important observation %
	underpinning the proof of \Cref{th:UrealU} below %
	is that $u\!\left(\iw\right)>-\infty$ (recall \Cref{obs:u>-infty}) along with the convention $+\infty-\infty\coloneqq-\infty$ leads to the equality %
		$u\!\left(\iw\right)=\sup_{\varphi\in\admR{\iw}}\inf_{S\in\mathscr{S}}\mathbb{E}_{\mathbb{P}}\!\left[U\!\left(W_{T}^{S}\!\left(\iw,\varphi\right)\right)\right]$, %
	where%
	\footnote{Hereafter,
		\begin{equation*}
			U^{\pm}\!\left(x\right)\coloneqq\max\!\left\{\pm U\!\left(x\right),0\right\},\quad\text{for all }x\in\dom\!\left(U\right)%
			,
		\end{equation*}
		denote the positive and negative parts of $U\!\left(\cdot\right)$, respectively.} %
	\begin{equation*}
		\admR{\iw}\coloneqq\bigcap_{S\in\mathscr{S}}\left\{\phi\in\mathscr{L}:~\mathbb{E}_{\mathbb{P}}\left[U^{-}\!\left(W^{S}_{T}\!\left(w_{0},\phi\right)\right)\right]<+\infty\right\}\neq\varnothing.%
	\end{equation*}
	This means that any investment leading to infinite disappointment in at least one possible model is automatically excluded from the robust portfolio problem as it can never be optimal, thereby effectively restricting the set of admissible controls from $\mathscr{L}$ to $\admR{\iw}$%
	.\footnote{We thank the referee for drawing our attention to this point.}%
\end{remark}

\begin{theorem}\label[th]{th:UrealU}
	Let $\iw\in\mathbb{R}$, and let \Cref{as:NA,as:U,as:wholerl%
	} hold. %
	Assume further that the following three conditions are all satisfied.
	\begin{enumerate}[label=\emph{(\roman*)}]
		\item
		$\mathscr{S}=\mathscr{S}^{\ast}$.
		
		\item
		There are $C>0$ and $\alpha\in\left[\left.0,1\right)\right.$ such that 
		\begin{equation}\label{maruszja}
			U\!\left(x\right)\leq C\left(x^{\alpha}+1\right)\ %
			\text{for all }x\geq0%
			.
		\end{equation}
		
		\item
		For all $S\in\mathscr{S}$,
		\begin{equation}\label{eq:lune}
			\frac{1}{\beta^S_t},\frac{1}{\kappa^S_t}
			,\norm{\Delta S_{t}}\in\mathscr{W}%
			\ \text{for all }t\in\left\{1,\ldots,T\right\}%
			,
		\end{equation}
		where $\beta^S_t$, $\kappa^S_t$ are the random variables given by \Cref{prop:NA}, and $\mathscr{W}$ is as in~\eqref{eq:W}. %
	\end{enumerate}		
	Then
	\ there exists $\opt\in\adm{\iw}$ such that
	\begin{equation*}
		u\!\left(\iw\right)%
		=\inf_{S\in\mathscr{S}}\mathbb{E}_{\mathbb{P}}\!\left[U\!\left(W_{T}^{S}\!\left(\iw,\opt\right)\right)\right]%
		<+\infty%
		. %
	\end{equation*}
\end{theorem}

\begin{proof}
	See~\Cref{app:Main}
	.
\end{proof}

\begin{remark}
	At this point, let us briefly look at the case of ``model-free'' utility maximisation; see e.g.\ \citet{acciaio2013}, where $\mathscr{S}$ consists of all the adapted price processes (or just the ones satisfying no arbitrage). A moment's reflection shows that, in those cases, only the identically $0$ strategy is admissible in the case of $U\!\left(\cdot\right)$ on the positive real axis. In the case of $U\!\left(\cdot\right)$ defined on the whole real line, every strategy except $0$ yields utility $-\infty$. In other words, the problem
becomes trivial.

	Another model-free approach has been outlined in \citet{bfm2016}. %
	However, it concentrates  uniquely on arbitrage, and probabilities do not appear in the natural way (i.e.\ their set $\mathscr{S}$ does not determine a set of probabilities). %
	In the special case $\mathscr{S}=\{\Omega\}$, the set of \emph{all} probabilities appears naturally, and we already addressed this case in the previous paragraph. For all other cases of \citet{bfm2016} we do not see even how the utility maximization problem could be properly formulated.
\end{remark}


\section{Examples}\label[sec]{sec:Example}


\begin{example}\label[ex]{toxun}
	When prices are assumed to be continuous-time It\^{o} processes, uncertainty in the drift and volatility are the classical setting for model robustness. This has been fervently studied, we refer to the seminal paper of \citet{epsteinji2014}. We sketch a simple discrete-time analogue below. %

	Consider the $t$-fold Cartesian product $\Omega_{t}\coloneqq\left\{-1,1\right\}^{t}$ for all $t\in\left\{0,1,\ldots,T\right\}$, and denote by $\pow{\Omega_{t}}$ the power set of $\Omega_{t}$. Set also $\Omega\coloneqq\Omega_T$. Define the mappings $\Pi_{t}:~\Omega
	\rightarrow\Omega_{t}$ as %
	\begin{equation*}
		\Pi_{t}\!\left(\omega\right)\coloneqq\left(\omega_{1},\ldots,\omega_{t}\right),\quad\text{for all }\omega=\left(\omega_{1},\ldots,\omega_{t},\ldots,\omega_{T}\right)\in\Omega
		,
	\end{equation*}
	and the filtration $\mathbb{F}$ as $\mathscr{F}_{t}\coloneqq\Pi_{t}^{-1}\!\left(\pow{\Omega_{t}}\right)$. %
	Fix $p\in\left(0,1\right)$, and equip the measurable space $\left(\Omega,2^\Omega\right)$ with the probability measure defined by %
	\begin{equation}\label{ex:probP}
		\mathbb{P}\!\left(\left\{\omega\right\}\right)\coloneqq\prod_{t=1}^{T}\left(p\,\delta_{1}\!\left(\left\{\proj_{t}\!\left(\omega\right)\right\}\right)+\left(1-p\right)\delta_{-1}\!\left(\left\{\proj_{t}\!\left(\omega\right)\right\}\right)\right),\quad\text{for all }\omega\in\Omega%
		,
	\end{equation}
	where $\delta_{x}$ is the Dirac measure at $x\in\mathbb{R}$,\footnote{%
	Given $x\in\mathbb{R}$, the \emph{Dirac measure} at $x$ is the probability measure $\delta_{x}:~\mathscr{B}\!\left(\mathbb{R}\right)\rightarrow\left\{0,1\right\}$ defined by
	\begin{equation*}
		\delta_{x}\!\left(B\right)\coloneqq\left\{
			\begin{array}{ll}
				1,& \text{if }x\in B,\\
				0,& \text{otherwise}.
			\end{array}
		\right.
	\end{equation*}
	} %
	and $\proj_{t}$ is the $t$-th projection map from $\Omega$ to $\left\{-1,1\right\}$.\footnote{Given a Cartesian product $X_{1}\times\ldots\times X_{n}$ and $k\in\left\{1,\ldots,n\right\}$, the $k$-th \emph{projection map} $\proj_{k}:~X_{1}\times\ldots\times X_{n}\rightarrow X_{k}$ is defined by
	\begin{equation*}
		\proj_{k}\!\left(x_{1},\ldots,x_{n}\right)\coloneqq x_{k},\quad\text{for all }\left(x_{1},\ldots,x_{n}\right)\in X_{1}\times\ldots\times X_{n}.
	\end{equation*}} %
	Moreover, let $s_{0}\in\mathbb{R}$ be given, and introduce the parameter space $\Theta\coloneqq\left[\ubar{\sigma},\bar{\sigma}\right]^{T}\times\mathbb{R}^{T}$ for some $0<\ubar{\sigma}<\bar{\sigma}$. %
	For every $\theta=\left(\sigma,\mu\right)\in\Theta$, define the real-valued process $S^{\theta}=\left\{S^{\theta}_{t}\right\}_{t\in\left\{0,1,\ldots,T\right\}}$ as
	\begin{align*}
		S^{\theta}_{t}\!\left(\omega\right)\coloneqq 
		s_{0}+\sum_{s=1}^{t}\left(\sigma\proj_{s}\!\left(\omega\right)+\mu\right)%
		,\quad\text{for all }\omega\in\Omega,\ t\in\left\{0,%
		1,\ldots,T\right\}.
	\end{align*}
	When $p=1/2$ then this model is a possible discrete-time analogue of the continuous-time Bachelier model. %
	
	Even though arbitrage opportunities exist for certain price processes, it is easily seen that each $S^{\theta}\in\mathscr{S}$ with $\left\vert\mu\right\vert<\sigma$ admits a unique equivalent martingale measure given by
	\begin{equation*}
		\mathbb{Q}^{\theta}\!\left(\left\{\omega\right\}\right)\coloneqq2^{-T}\prod_{t=1}^{T}\left(1-\proj_{t}\!\left(\omega\right)\frac{\mu}{\sigma}\right),\quad\text{for all }\omega\in\Omega.
	\end{equation*}
	In addition, $S^{\theta}$ has independent increments for every $\theta\in\Theta$, which entails that condition~\eqref{eq:intW} holds and \eqref{eq:lune} is satisfied by any arbitrage-free process in $\mathscr{S}$, as observed by \citet[Proposition~7.1]{rs05}. %
	Note also that $\mathscr{S}$ may contain price processes whose laws are mutually singular. %
	
	Consider now $\mathscr{P}\coloneqq\left\{\Law\!\left(S\right):~S\in\mathscr{S}\right\}$, which is a subset of the space of all probability measures on $\left(\mathbb{R},\mathscr{B}\!\left(\mathbb{R}\right)\right)$. This simple model class is a typical example of model-misspecification where the ``volatility'' parameter $\sigma$ and the drift parameter $\mu$ are unknown. It can be treated in our framework. %
	It does not, however, satisfy the condition on the sets of probability measures $\mathscr{P}_{t}\!\left(\omega\right)$ in \citet[Section~3]{nutz2016} or \citet[Section~2]{bartl2017}; namely, the family of possible models for the stock price at time $t+1$ given the state $\omega$ at time $t$ is not convex (as convexity here would mean being closed under taking mixtures of probabilities). %

	One could consider various generalizations of this example: $\mu_{t}$ and $\sigma_{t}$ could be time-dependent and even $\mathscr{F}_{t-1}$-measurable; taking $\Omega:=\mathbb{R}^{T}$, we could consider more general noise processes with possibly continuous laws, etc.
\end{example}

\begin{example}
	Most papers in the standard multiple-priors framework (e.g.\ \citet{nutz2016,bouchardnutz2015,bartl2017}) assume the ``time-consistency'' property. It is possible to construct an example where the corresponding $\mathscr{P}$ fails to satisfy this property, but can be treated in our framework. Recall also \citet{bck2019}, who could drop time-consistency using additional set-theoretical axioms. %
	
	Fix $T\coloneqq 2$ and $d=1$ (i.e., consider a single risky asset). Take $\Omega_{1}=\mathbb{R}$ and $\Omega\coloneqq\Omega_{2}=\mathbb{R}^{2}$. %
	Let $\varepsilon_{i}$, $i\in\left\{1,2\right\}$ be two independent and identically distributed random variables with $\mathbb{P}\!\left\{\varepsilon_{i}=\pm 1\right\}=1/2$, and define the filtration $\mathbb{F}$ such that $\mathscr{F}_{0}$ is the $\sigma$-algebra of $\mathbb{P}$-zero sets, $\mathscr{F}_{1}\coloneqq\sigma\!\left(\varepsilon_{1}\right)$, and $\mathscr{F}_{2}\coloneqq\sigma\!\left(\varepsilon_{1},\varepsilon_{2}\right)$. %
	
	Let $s_{0}=0$, and for each $\theta=\left(\mu_{1},\mu_{2}\right)\in\mathbb{R}^{2}$ define the real-valued price process $S^{\theta}=\left\{S_{t}^{\theta}\right\}_{t\in\left\{0,1,2\right\}}$ as %
	\begin{equation*}
		S_{t}^{\theta}\coloneqq s_{0}+\sum_{s=1}^{t}\left(\mu_{s}+\varepsilon_{s}\right),\quad\text{for all }t\in\left\{0,1,2\right\},
	\end{equation*}
	Assume further that there are only two possible price models; namely, $\mathscr{S}=\left\{S^{\theta}:~\theta\in\Theta\right\}$, where $\Theta=\left\{\theta_{1}=\left(0.1,0.3\right),\theta_{2}=\left(0.2,0.5\right)\right\}$. Note that $\mathscr{S}=\NAset$. %
	Lastly, introduce the notation $\mathscr{P}\coloneqq\left\{\Law\!\left(S\right):~S\in\mathscr{S}\right\}$. %
	
	Time-consistency means that there is a collection of probabilities $\mathscr{P}_{0}$ on $\mathbb{R}$ and a set-valued mapping $\mathscr{P}_{1}$ from $\mathbb{R}$ to the set of probability measures on $\mathbb{R}$ such that $\mathscr{P}$ consists of all probability measures $\mathbb{P}$ on $\mathbb{R}^{2}$ admitting the decomposition%
	\footnote{Here, $\mathscr{B}\!\left(\mathbb{R}^{2}\right)$ denotes the Borel $\sigma$-algebra of $\mathbb{R}^{2}$ (i.e., the smallest $\sigma$-algebra containing all open subsets of $\mathbb{R}^{2}$).}
	\begin{equation*}
		\mathbb{P}\!\left(A\right)=\int_{\mathbb{R}}\int_{\mathbb{R}}\bbOne_{A}\!\left(\omega_{1},\omega_{2}\right)\,P_{1}\!\left(\omega_{1},d\omega_{2}\right)\mathbb{P}_{0}\!\left(d\omega_{1}\right)\text{ for all }A\in\mathscr{B}\!\left(\mathbb{R}^{2}\right)
	\end{equation*}
	($\mathbb{P}=P_{0}\otimes P_{1}$ in abbreviated form), for some $P_{0}\in\mathscr{P}_{0}$ and $P_{1}$ satisfying $P_{1}\!\left(\omega\right)\in\mathscr{P}_{1}\!\left(\omega\right)$ for all $\omega\in\mathbb{R}$; see \citet{bouchardnutz2015}. %
	
	Next, observe that, for every $\theta=\left(\mu_{1},\mu_{2}\right)\in\mathbb{R}^{2}$, each of the stock prices $S^{\theta}_{1}$ and $S^{\theta}_{2}$ has discrete law with support $\left\{s_{0}+\mu_{1}-1,s_{0}+\mu_{1}+1\right\}$ and $\left\{s_{0}+\mu_{1}+\mu_{2}-2,s_{0}+\mu_{1}+\mu_{2},s_{0}+\mu_{1}+\mu_{2}+2\right\}$, respectively; moreover, 
	\begin{alignat*}{3}
		&\mathbb{P}\!\left\{S^{\theta}_{1}=s_{0}+\mu_{1}\pm1\right\}=\frac{1}{2},\qquad &&\\
		&\mathbb{P}\!\left\{S^{\theta}_{2}=s_{0}+\mu_{1}+\mu_{2}\pm2\right\}=\frac{1}{4},\qquad &&\mathbb{P}\!\left\{S^{\theta}_{2}=s_{0}+\mu_{1}+\mu_{2}\right\}=\frac{1}{2}.
	\end{alignat*}
	
	Assume, with a view to a contradiction, that time-consistency holds. Denoting by $\mathbb{P}_{i}$ the probability law of $S^{\theta_{i}}$ for $i\in\left\{1,2\right\}$, straightforward computations yield
	\begin{align*}
		&\mathbb{P}_{1}=P_{0}^{1}\otimes P_{1}^{1},\\
		&\mathbb{P}_{2}=P_{0}^{2}\otimes P_{1}^{2},
	\end{align*}
	where $P_{0}^{1}\in\mathscr{P}_{0}$ is the discrete distribution with $P_{0}^{1}\!\left\{0.1\pm1\right\}=1/2$; %
	$P_{1}^{1}$ is such that $P_{1}^{1}\!\left(0.1\pm1\right)\in\mathscr{P}_{1}\!\left(0.1\pm1\right)$ are the discrete distributions with
	\begin{align*}
		&P_{1}^{1}\!\left(0.1+1,\left\{0.1+0.3+2\right\}\right)=P_{1}^{1}\!\left(0.1+1,\left\{0.1+0.3\right\}\right)=1/2,\\%
		&P_{1}^{1}\!\left(0.1-1,\left\{0.1+0.3-2\right\}\right)=P_{1}^{1}\!\left(0.1-1,\left\{0.1+0.3\right\}\right)=1/2%
		; %
	\end{align*}
	$P_{0}^{2}\in\mathscr{P}_{0}$ is the discrete distribution with $P_{0}^{2}\!\left\{0.2\pm1\right\}=1/2$; and $P_{1}^{2}$ is such that $P_{1}^{2}\!\left(0.2\pm1\right)\in\mathscr{P}_{1}\!\left(0.2\pm1\right)$ are the discrete distributions with
	\begin{align*}
		&P_{1}^{2}\!\left(0.2+1,\left\{0.2+0.5+2\right\}\right)=P_{1}^{2}\!\left(0.2+1,\left\{0.2+0.5\right\}\right)=1/2,\\
		&P_{1}^{2}\!\left(0.2-1,\left\{0.2+0.5-2\right\}\right)=P_{1}^{2}\!\left(0.2-1,\left\{0.2+0.5\right\}\right)=1/2%
		. %
	\end{align*}
	However, $\bar{\mathbb{P}}\coloneqq P_{0}^{1}\otimes P_{1}^{2}$ is not an element of $\mathscr{P}$, a contradiction. %
	A different example can be found in \citet[Appendix~D, Example~3]{riedel09}. %
\end{example}

\begin{example}\label[ex]{horus}
	This is a much simplified version of Example~2.7 of \citet{bartl2017}.
	Let us fix $s_{0}\in\mathbb{R}$ and $\sigma>0$, set
	\begin{equation*}
		S_{t}\!\left(\omega\right)\coloneqq 
		s_{0}+\sigma\sum_{s=1}^{t}\proj_{s}\!\left(\omega\right)%
		,\quad\text{for all }\omega\in\Omega,\ t\in\left\{0,%
		1,\ldots,T\right\},
	\end{equation*}
	and define $\mathscr{P}\coloneqq\left\{\mathbb{P}^{p}:~p\in\left(0,1\right)\right\}$ where each $\mathbb{P}^{p}$ is of the form~\eqref{ex:probP}. While $\mathscr{P}$ satisfies the conditions of \citet{nutz2016}, we cannot incorporate this example into our framework in a natural way. Indeed, one would need an information filtration $\mathbb{F}=\left\{\mathscr{F}_{t}\right\}_{t\in\left\{0,1,\ldots,T\right\}}$ such that	the price increments (which can be constant times Bernoulli with arbitrary parameters) are measurable with respect to it. This is feasible only if $\mathbb{F}$ is generated by continuous random variables. This is, however, unnatural, as the investment decision at time $t+1$ should be measurable with respect to the price movements up to $t$, which are \emph{discrete-valued} random variables. %
	Hence there is no way of incorporating this example into our framework in a	meaningful way. %
\end{example}

\Cref{toxun,horus} show that, while the usual approach is capable of capturing uncertainty at the level of the ``probabilistic skeleton'' of the process (e.g.\ probabilities on a tree), our approach can be successful for various parametrisations when the filtration (the information structure) is kept fixed.

Below is another, ``non-parametric'' example that can be covered by our method.

\begin{example}
	On a fixed probability space $\left(\Omega,\mathscr{F},\mathbb{P}\right)$, consider a collection $\left\{\varepsilon_{1},\ldots,\varepsilon_{T}\right\}$ of independent and standard uniform random variables. Define the filtration $\mathbb{F}$ as $\mathscr{F}_{t}\coloneqq\sigma\left(\varepsilon_{1},\ldots,\varepsilon_{t}\right)$ for all $t\in\left\{0,1,\ldots,T\right\}$, let $s_{0}\in\mathbb{R}$ be given, and denote by $\Theta$ the set of $T$-tuples of $\mathscr{F}_{t-1}\otimes\mathscr{B}\!\left(\left[0,1\right]\right)$-measurable functions $\theta_{t}:~\Omega\times\left[0,1\right]\rightarrow\mathbb{R}$, $t\in\left\{1,\ldots,T\right\}$ that are non-decreasing in their second variable. %

	For every $\theta=(\theta_{1},\ldots,\theta_{T})\in\Theta$, let $S^{\theta}=\{S^{\theta}_{t}\}_{t\in\left\{0,1,\ldots,T\right\}}$ be the real-valued process given by
	\begin{equation*}
		S^{\theta}_{t}\!\left(\omega\right)\coloneqq s_{0}+\sum_{s=1}^{t}\theta_{s}\!\left(\omega,\varepsilon_{s}\!\left(\omega\right)\right),\quad\text{for all }\omega\in\Omega,\ t\in\left\{0,%
		1,\ldots,T\right\}.
	\end{equation*}
	Let now $\mathscr{S}\subseteq\left\{S^{\theta}:\ \theta\in\Theta\right\}$. If $\mathscr{S}$ contains $S^{\theta^{\ast}}$ for some $\theta^{\ast}$ such that $\theta^{\ast}_{t}\!\left(\omega,\cdot\right)$ takes on positive as well as negative values with positive Lebesgue measure for all $t$ and $\omega$, then $S^{\theta^{\ast}}$	clearly satisfies the no-arbitrage assumption as well as \eqref{eq:nonred}. Notice that an arbitrary probability law on $\mathbb{R}$ can be represented by a random variable that is a non-decreasing function of a standard uniform random variable.

	We now show that this example subsumes Example~2.8 of \citet{bartl2017}, except that we prefer to use an additive dynamics instead of the multiplicative one of \citet{bartl2017}. %
	For every $t\in\left\{1,\ldots,T\right\}$, fix $\mathscr{F}_{t-1}$-measurable real-valued random variables $\underline{\mu}_{t}\leq \overline{\mu}_{t}$, and positive $\mathscr{F}_{t-1}$-measurable random variables $\sigma_{t},\epsilon_{t}$. Now we can define $\mathscr{S}$ as the set of price models $S=\left\{S_{t}\right\}_{t\in\left\{0,1,\ldots,T\right\}}$ satisfying $S_{0}\coloneqq s_{0}\in\mathbb{R}$ and
	\begin{equation*}
		S_{t}=S_{t-1}+\mu_{t}+Y_{t},\quad\text{for all }t\in\left\{1,\ldots,T\right\},
	\end{equation*}
	where $\mu_{t}$ ranges over $\left[\underline{\mu}_{t},\overline{\mu}_{t}\right]$-valued $\mathscr{F}_{t-1}$-measurable random variables, and $Y_{t}$ ranges over the random variables whose conditional law with respect to $\mathscr{F}_{t-1}$ is $\epsilon_{t}\!\left(\omega\right)$-close to the Gaussian law with mean $0$ and variance $\sigma_{t}\!\left(\omega\right)$, for almost every $\omega$, in some suitable distance. %
	We do not enter into further technical details.
\end{example}

\begin{example}[Option trading]
	Let $Z_{t}$, $t\in\left\{1,\ldots,T\right\}$ be $\mathbb{R}^{m}$-valued random variables on some probability space $\left(\Omega,\mathscr{F},\mathbb{P}\right)$, representing the prices of $m$ stocks, and let $Z_{0}\in\mathbb{R}^{m}$ be deterministic. Let $\mathscr{F}_{0}$ be the set of $\mathbb{P}$-zero sets, and let
	\begin{equation*}
		\mathscr{F}_{t}\coloneqq\mathscr{F}_{0}\vee\sigma\!\left(Z_{s},\ s\in\left\{1,\ldots,t\right\}\right)
	\end{equation*}
	for $t\in\left\{1,\ldots,T\right\}$. %
	We assume that \hyperlink{eq:NAS}{NA($Z$)} holds, hence there is at least one probability, equivalent to $\mathbb{P}$, under which $Z$ is a martingale. Denote by $\mathscr{Q}$ the set of all such probabilities. %
	Let $G$ be an $\mathbb{R}^{d}$-valued $\mathscr{F}_{T}$-measurable random variable, representing the terminal value of $d$ options written on these stocks, and let $g\in\mathbb{R}^{d}$ be a vector representing the (known) prices of these options at time $0$.

	Assume that $\mathscr{Q}\!\left(g\right)\coloneqq\left\{\mathbb{Q}\in\mathscr{Q}:~\mathbb{E}_{\mathbb{Q}}\!\left[G\right]=g\right\}\neq\varnothing$, and
	fix some non-empty $\mathscr{R}\subset\mathscr{Q}\!\left(g\right)$. Define
	\begin{equation*}
		\mathscr{S}\coloneqq\left\{\mathbb{E}_{\mathbb{Q}}\!\left[\left.G\right\vert\mathscr{F}_{t}\right],\ t\in\left\{0,\ldots,T\right\}:\ \mathbb{Q}\in\mathscr{R}%
		\right\},
	\end{equation*}
	the collection of possible option price processes under the various risk-neutral probabilities $\mathbb{Q}\in\mathscr{R}$.

	Results of our paper apply to the above setting where there is uncertainty about the pricing measure and the investor with worst-case robust preferences tries to optimally trade in these options.
\end{example}


\section{Conclusion}\label[sec]{sec:Conc}


This paper provides a positive answer to the question of existence of solutions to the worst-case robust utility maximisation problem in an alternative framework for model uncertainty. %
It is desirable to extend the arguments of this work to continuous-time models, as done in the companion paper of \citet{chaurasonyi2018} in a market where proportional transaction costs are present. %

It is also worth investigating whether there exists a price process $S_{0}\in\mathscr{S}$ with respect to which the standard utility maximisation problem is equivalent to the robust utility maximisation problem (by analogy with e.g.\ the least favourable measure in \citet{schied2005}).

The fundamental questions of arbitrage and superhedging have been intensively studied in the presence of multiple priors, see \Cref{obs:RNA} above, as well as \cite{acciaio2013,bouchardnutz2015,bfm2016,bfhmo2018} and the references therein. These questions have not yet been addressed in the setting of the present work, and will be the subject of future research.


\FloatBarrier

\appendix
\section{Appendix: Proofs}\label[app]{app:Main}

\setlist[enumerate,1]{
  leftmargin=*,
}
\setlist[enumerate,2]{
	wide=0pt
}


This appendix contains the proofs of the results presented in \Cref{sec:Main}. %



\paragraph{Proof of \Cref{lem:key}}\pprangestart{proof:key}%
Let $S^{\ast}\in\NAset\neq\varnothing$ be given. %
We construct the random variables $G^{S^{\ast}}_{1},\ldots,G^{S^{\ast}}_{T}$ recursively as follows. %
\begin{enumerate}[label=\emph{(\roman*)},wide,labelindent=0pt,labelwidth=!
]

	\item
	Define $G^{S^{\ast}}_{1}\coloneqq\iw/\beta^{S^{\ast}}_{1}$, where $\beta^{S^{\ast}}_{1}$ is given by \Cref{prop:NA}, 
	and let $\phi\in%
	\mathscr{L}\!\left(\iw,S^{\ast}\right)%
	$. %
	Clearly, $G^{S^{\ast}}_{1}$ is both strictly positive and finite on the $\mathbb{P}$-full set $\left\{\beta^{S^{\ast}}_{1}>0\right\}$. %
	It follows from $\iw+\innerp{\hat{\phi}^{S^{\ast}}_{1}}{\Delta S^{\ast}_{1}}=\iw+\innerp{\phi_{1}}{\Delta S^{\ast}_{1}}=W^{S^{\ast}}_{1}\!\left(\iw,\phi\right)\geq0$ $\mathbb{P}$-a.s.\ %
	(recall \Cref{obs:proj}) %
	that
	\begin{equation*}
		\mathbb{P}\!\left(\left\{\norm{
		\hat{\phi}^{S^{\ast}}_{1}%
		}>G^{S^{\ast}}_{1}\right\}\cap\left\{\innerp{
		\hat{\phi}^{S^{\ast}}_{1}%
		}{\Delta S^{\ast}_{1}}\leq-\beta^{S^{\ast}}_{1}\norm{
		\hat{\phi}^{S^{\ast}}_{1}%
		}\right\}\right)%
		\leq\mathbb{P}\!\left\{\innerp{
		\hat{\phi}^{S^{\ast}}_{1}%
		}{\Delta S^{\ast}_{1}}<-\iw\right\}=0
		.
	\end{equation*}
	This inequality %
	together with \eqref{eq:altNA} %
	as well as the $\mathscr{F}_{0}$-measurability of both $
	\hat{\phi}^{S^{\ast}}_{1}%
	$ and $G^{S^{\ast}}_{1}$ %
	yields%
	\footnote{Here, $\bbOne_{A}:~X\rightarrow\left\{0,1\right\}$ is the \emph{indicator} function of the set $A\subseteq X$, defined as
	\begin{equation*}
		\bbOne_{A}\!\left(x\right)\coloneqq\left\{
			\begin{array}{ll}
				1,& \text{if }x\in A,\\
				0,& \text{otherwise}.
			\end{array}
		\right.
	\end{equation*}
	} %
	\begin{equation*}
		\mathbb{E}_{\mathbb{P}}\!\left[\kappa^{S^{\ast}}_{1}\bbOne_{\left\{\norm{
		\hat{\phi}^{S^{\ast}}_{1}%
		}>G^{S^{\ast}}_{1}\right\}}\right]\leq\mathbb{P}\!\left(
		\left\{\norm{
		\hat{\phi}^{S^{\ast}}_{1}%
		}>G^{S^{\ast}}_{1}\right\}\cap\left\{\innerp{
		\hat{\phi}^{S^{\ast}}_{1}%
		}{\Delta S^{\ast}_{1}}\leq-\beta^{S^{\ast}}_{1}\norm{
		\hat{\phi}^{S^{\ast}}_{1}%
		}\right\}
		\right)=0\ \mathbb{P}\text{-a.s.}%
		.
	\end{equation*}
	Given that $\kappa^{S^{\ast}}_{1}>0$ $\mathbb{P}$-a.s., we conclude %
	$\mathbb{P}\!\left\{\norm{
	\hat{\phi}^{S^{\ast}}_{1}%
	}>G^{S^{\ast}}_{1}\right\}=0$. %

	\item
	Suppose that, for some $t\in\left\{1,\ldots,T\right\}$, we have obtained $G^{S^{\ast}}_{1},\ldots,G^{S^{\ast}}_{t}$ satisfying the conditions of \Cref{lem:key}. Setting $G^{S^{\ast}}_{t+1}\coloneqq\left(\iw+\sum_{s=1}^{t}G^{S^{\ast}}_{s}\norm{\Delta S^{\ast}_{s}}\right)/\beta^{S^{\ast}}_{t+1}$%
	, its measurability, $\mathbb{P}$-a.s.\ strict positivity and finiteness properties are straightforward. Moreover, given any %
	$\phi\in\mathscr{L}\!\left(\iw,S^{\ast}\right)$, we can use the Cauchy-Schwarz inequality %
	(along with \Cref{obs:proj} again) %
	to obtain
	\begin{equation*}
		0\leq W^{S^{\ast}}_{t+1}\!\left(\iw,\phi\right)%
		\leq\iw+\sum_{s=1}^{t}\norm{
		\hat{\phi}^{S^{\ast}}_{s}%
		}\norm{\Delta S^{\ast}_{s}}+\innerp{
		\hat{\phi}^{S^{\ast}}_{t+1}%
		}{\Delta S^{\ast}_{t+1}}\leq\beta^{S^{\ast}}_{t+1}G^{S^{\ast}}_{t+1}+\innerp{
		\hat{\phi}^{S^{\ast}}_{t+1}%
		}{\Delta S^{\ast}_{t+1}}\ \mathbb{P}\text{-a.s.}%
		.
	\end{equation*}
	As a consequence,
	\begin{multline*}
		\mathbb{P}\!\left(\left\{\norm{
		\hat{\phi}^{S^{\ast}}_{t+1}%
		}>G^{S^{\ast}}_{t+1}\right\}\cap\left\{\innerp{
		\hat{\phi}^{S^{\ast}}_{t+1}%
		}{\Delta S^{\ast}_{t+1}}\leq-\beta^{S^{\ast}}_{t+1}\norm{
		\hat{\phi}^{S^{\ast}}_{t+1}%
		}\right\}\right)
		\leq\mathbb{P}\!\left\{\innerp{
		\hat{\phi}^{S^{\ast}}_{t+1}%
		}{\Delta S^{\ast}_{t+1}}<-\beta^{S^{\ast}}_{t+1}G^{S^{\ast}}_{t+1}\right\}=0,
	\end{multline*}
	and a similar argument to that of the previous step leads to 
	$\norm{
	\hat{\phi}^{S^{\ast}}_{t+1}%
	}\leq G^{S^{\ast}}_{t+1}$ $\mathbb{P}$-a.s.%
	.%
	\qed
	
\end{enumerate}

\pprangeend{proof:key}


\paragraph{Proof of \Cref{th:Ubounded}}\pprangestart{proof:Ubounded}%
Well-posedness %
of the optimisation problem~\eqref{eq:rEU} %
is trivial, as
\begin{equation*}
	\mathbb{E}_{\mathbb{P}}\!\left[U^{+}\!\left(W^{S}_{T}\!\left(\iw,\varphi\right)\right)\right]
	\leq U\!\left(+\infty\right)
\end{equation*}
for all $\varphi\in\adm{\iw}$%
\ and $S\in\mathscr{S}$%
. The 
proof %
of existence of an optimal portfolio %
is organised into three parts.
\begin{enumerate}[label=\emph{(\roman*)},wide,labelindent=0pt,labelwidth=!
]

	\item
	We claim that, for every $S\in\mathscr{S}$, the functional $\Xi^{S}:~\adm{\iw}\rightarrow\mathbb{R}\cup\!\left\{-\infty\right\}%
	$ %
	defined by
	\begin{equation*}
		\Xi^{S}\!\left(\varphi\right)\coloneqq\mathbb{E}_{\mathbb{P}}\!\left[U\!\left(W_{T}^{S}\!\left(\iw,\varphi\right)\right)\right],\quad\text{for all }\varphi\in\adm{\iw},
	\end{equation*}
	is sequentially upper semicontinuous. To see this, let $\varphi\in\adm{\iw}$ 
	and consider a sequence $\seq{\varphi^{n}}\subseteq\adm{\iw}$ converging to $\varphi$%
	\ with respect to $\tau$. Then $\seq{\varphi^{n}_{t}}$ converges in probability to $\varphi_{t}$ for all $t\in\left\{1,\ldots,T\right\}$, which combined with the continuous mapping theorem gives the convergence in probability of the sequence of random variables %
	$\seq{U^{\pm}\!\left(W^{S}_{T}\!\left(\iw,\varphi^{n}\right)\right)}$ to $U^{\pm}\!\left(W^{S}_{T}\!\left(\iw,\varphi\right)\right)$. Hence,
	\begin{align*}
		\limsup_{n\rightarrow+\infty}\Xi^{S}\!\left(\varphi^{n}\right)&\leq\limsup_{n\rightarrow+\infty}\mathbb{E}_{\mathbb{P}}\!\left[U^{+}\!\left(W_{T}^{S}\!\left(\iw,\varphi^{n}\right)\right)\right]-\liminf_{n\rightarrow+\infty}\mathbb{E}_{\mathbb{P}}\!\left[U^{-}\!\left(W_{T}^{S}\!\left(\iw,\varphi^{n}\right)\right)\right]\\
		&\leq\mathbb{E}_{\mathbb{P}}\!\left[U^{+}\!\left(W_{T}^{S}\!\left(\iw,\varphi\right)\right)\right]-\mathbb{E}_{\mathbb{P}}\!\left[U^{-}\!\left(W_{T}^{S}\!\left(\iw,\varphi\right)\right)\right]=\Xi^{S}\!\left(\varphi\right),
	\end{align*}
	where the second inequality is a consequence of Fatou's lemma and the reverse Fatou lemma (whose use is justified by $U^{+}\!\left(W_{T}^{S}\!\left(\iw,\varphi^{n}\right)\right)\leq U\!\left(+\infty\right)$ for all $n\in\mathbb{N}$). %
	
	Now note that $\mathscr{L}$ is first-countable, since any finite product of metrisable spaces is metrisable and every metrisable space is first-countable. But this implies that the subspace $\adm{\iw}$ is also first-countable. In addition, recall that sequential upper semicontinuity and upper semicontinuity are equivalent notions in first-countable spaces. Consequently, the functional $\Xi:~\adm{\iw}\rightarrow\mathbb{R}\cup\!\left\{-\infty\right\}%
	$ defined as
	\begin{equation*}
		\Xi\!\left(\varphi\right)\coloneqq\inf_{S\in\mathscr{S}}\mathbb{E}_{\mathbb{P}}\!\left[U\!\left(W_{T}^{S}\!\left(\iw,\varphi\right)\right)\right]%
		,\quad\text{for all }\varphi\in\adm{\iw},
	\end{equation*}
	being the pointwise infimum of the non-empty collection $\{\Xi^{S}\!\left(\cdot\right)\}_{S\in\mathscr{S}}$ of upper semicontinuous functions, is itself upper semicontinuous.

	\item\label{item:maxseq}
	Consider a 
	sequence $\seq{\varphi^{n}}\subseteq\adm{\iw}$ 
	such that, for all $n\in\mathbb{N}$,
	\begin{equation}\label{eq:maxseq}
		\Xi\!\left(\varphi^{n}\right)=\inf_{S\in\mathscr{S}}\mathbb{E}_{\mathbb{P}}\!\left[U\!\left(W^{S}_{T}\!\left(\iw,\varphi^{n}\right)\right)\right]>u\!\left(\iw\right)-\frac{1}{n}%
		.
	\end{equation}
	By virtue of \Cref{as:NA}, we can find some process $S^{\ast}$ in $\NAset$ for the risky asset prices. %
	From now on we will write $\varphi^{n}_{t}$ instead of the projections $\hat{\varphi}^{S^{\ast},n}_{t}$,
	for simplicity's sake.
	
	Then $\seq{
	\varphi^{n}_{1}%
	}$ is a sequence in $\Lspace[d]{0}{}{\mathscr{F}_{0}}$ with $\sup_{n\in\mathbb{N}}\norm{
	\varphi^{n}_{1}%
	}\leq G^{S^{\ast}}_{1}<+\infty$ $\mathbb{P}$-a.s.\ by \Cref{lem:key}, so it follows from \citet[Lemma~3.2]{schachermayer92} that there exists a sequence $\seq{\theta^{n}_{1}}$ of convex combinations of the tail of $\seq{
	\varphi^{n}_{1}%
	}$ (i.e., $\theta^{n}_{1}\in\conv\!\left\{
	\varphi^{m}_{1}%
	:~m\geq n\right\}$ for all $n\in\mathbb{N}$) converging in probability to some 
	$\opt_{1}\in\Lspace[d]{0}{}{\mathscr{F}_{0}}$.

	Next, denote by $\seq{\theta^{n}_{2}}$ the sequence of the same convex combinations of $\seq{
	\varphi^{n}_{2}%
	}$, and note that $\sup_{n\in\mathbb{N}}\norm{\theta^{n}_{2}}\leq G^{S^{\ast}}_{2}<+\infty$ $\mathbb{P}$-a.s. 
	Applying Lemma~3.2 in \citet{schachermayer92} again, this time to $\seq{\theta^{n}_{2}}$, 
	yields a sequence $\seq{\vartheta^{n}_{2}}$ of elements 
	$\vartheta^{n}_{2}\in\conv\!\left\{\theta^{m}_{2}:~m\geq n\right\}$ 
	(thus, of convex combinations of the tail of $\seq{
	\varphi^{n}_{2}%
	}$) that converges in probability to some $\opt_{2}\in\Lspace[d]{0}{}{\mathscr{F}_{1}}$. %
	Hence, letting $\seq{\vartheta^{n}_{1}}$ be given by the same convex combinations (of $\seq{
	\varphi^{n}_{1}%
	}$) as $\seq{\vartheta^{n}_{2}}$ (of $\seq{
	\varphi^{n}_{2}%
	}$), it is immediate that not only $\seq{\vartheta^{n}_{1}}$ converges in probability to 
	$\opt_{1}$, but also $\left(\vartheta^{n}_{1},\vartheta^{n}_{2}\right)\in\conv\!\left\{\left(
	\varphi^{m}_{1},\varphi^{m}_{2}%
	\right):~m\geq n\right\}$ for all $n\in\mathbb{N}$.

	Proceeding recursively in this way, we construct in a finite number of steps a process $\opt\in\mathscr{L}$ and a sequence $\seq{\phi^{n}}$ such that $\phi^{n}\in\conv\!\left\{
	\varphi^{n},\varphi^{n+1}%
	,\ldots\right\}$ for all $n\in\mathbb{N}$, and convergence in probability of $\seq{\phi^{n}_{t}}$ to $\opt_{t}$ holds for all $t\in\left\{1,\ldots,T\right\}$.
	
	Lastly, given any $n\in\mathbb{N}$, observe that $\phi^{n}_{t}$ belongs to $D_{t}^{S^{\ast}}$ $\mathbb{P}$-a.s.\ for all $t\in\left\{1,\ldots,T\right\}$; moreover, we can easily check that $W^{S}\!\left(\iw,\phi^{n}\right)\in\conv\!\left\{W^{S}\!\left(\iw,\varphi^{m}\right):~m\geq n\right\}$ for all $S\in\mathscr{S}$, provided that \eqref{eq:nonred} holds. %

	\item\label{item:opt}
	It remains only to verify that the process $\opt\in\mathscr{L}$ found in \ref{item:maxseq} is a
	\ solution of \eqref{eq:rEU}. That $\opt$ is an admissible portfolio is straightforward by the convexity and closedness properties of $\adm{\iw}$. Furthermore, we use that $U\!\left(\cdot\right)$ is concave and inequality~\eqref{eq:maxseq} to see that %
	\begin{equation*}
		\Xi\!\left(\phi^{n}\right)=%
		\inf_{S\in\mathscr{S}}\mathbb{E}_{\mathbb{P}}\!\left[U\!\left(W^{S}_{T}\!\left(\iw,\phi^{n}\right)\right)\right]\geq u\!\left(\iw\right)-\frac{1}{n}
	\end{equation*}
	for all $n\in\mathbb{N}$
	, which %
	%
	together with (sequential) %
	upper semicontinuity 
	yields %
	$\Xi\!\left(\opt\right)\geq\limsup_{n\rightarrow+\infty}\Xi\!\left(\phi^{n}\right)\geq u\!\left(\iw\right)$%
	.%
	\qed
	
\end{enumerate}

\pprangeend{proof:Ubounded}


\paragraph{Proof of \Cref{th:intW}}\pprangestart{proof:intW}%
A quick inspection of the proof of \Cref{lem:key} reveals that, under condition~\eqref{eq:intW},%
\ the random variables $G^{S^{\ast}}_{1},\ldots,G^{S^{\ast}}_{T}$ %
also %
belong to $\mathscr{W}$%
. %
The proof proceeds in three steps.
\begin{enumerate}[label=\emph{(\roman*)},wide,labelindent=0pt,labelwidth=!
]

	\item
	To show that \eqref{eq:rEU} is well-posed, first %
	observe that, %
	because $U\!\left(\cdot\right)$ is concave, 
	there is $C>0$ such that $U\!\left(x\right)\leq C\left(\left\vert x\right\vert+1\right)$ for all $x>0$
	. Using %
	this linear growth condition, 
	the Cauchy-Schwarz inequality, %
	and \Cref{lem:key}%
	\ (while also recalling \Cref{obs:proj})%
	, %
	\begin{equation}\label{eq:revFatouSast}
		U\!\left(W^{S
		}_{T}\!\left(\iw,\varphi\right)\right)%
		=U\!\left(W^{S
		}_{T}\!\left(\iw,\hat{\varphi}^{S^{\ast}}\right)\right)%
		\leq C\left(1+\iw+\sum_{s=1}^{T}G^{S^{\ast}}_{s}\norm{\Delta S
		_{s}}\right)
	\end{equation}
	for all $\varphi\in\adm{\iw}$%
	\ and $S\in\mathscr{S}$%
	, thus %
		$u\!\left(\iw\right)\leq C\left(1+\iw+\sum_{s=1}^{T}\mathbb{E}_{\mathbb{P}}\!\left[\left(G^{S^{\ast}}_{s}\right)^{2}\right]^{\frac{1}{2}}\mathbb{E}_{\mathbb{P}}\!\left[\norm{\Delta S^{\ast}_{s}}^{2}\right]^{\frac{1}{2}}\right)<+\infty$.

	\item
	Let the functionals $\Xi^{S}\!\left(\cdot\right)$ for every $S\in\mathscr{S}$, and $\Xi\!\left(\cdot\right)$ be as in the proof of \Cref{th:Ubounded}. To establish upper semicontinuity of $\Xi\!\left(\cdot\right)$, it suffices to show that $\Xi^{S}\!\left(\cdot\right)$ is sequentially upper semicontinuous for all $S\in\mathscr{S}$, so fix $\varphi\in\adm{\iw}$ and an arbitrary sequence $\seq{\varphi^{n}}\subseteq\adm{\iw}$ convergent to $\varphi$ in the topology $\tau$. Since $\seq{U^{\pm}\!\left(W^{S}_{T}\!\left(\iw,\varphi^{n}\right)\right)}$ converge in probability to $U^{\pm}\!\left(W^{S}_{T}\!\left(\iw,\varphi\right)\right)$, and $\seq{U^{+}\!\left(W^{S}_{T}\!\left(\iw,\varphi^{n}\right)\right)}$ is dominated by the 
	integrable
	\ random variable on the right-hand side of~\eqref{eq:revFatouSast}, we can apply both Fatou's lemma and the reverse Fatou lemma to obtain
	\begin{align*}
		\limsup_{n\rightarrow+\infty}\Xi^{S}\!\left(\varphi^{n}\right)&\leq\limsup_{n\rightarrow+\infty}\mathbb{E}_{\mathbb{P}}\!\left[U^{+}\!\left(W_{T}^{S}\!\left(\iw,\varphi^{n}\right)\right)\right]-\liminf_{n\rightarrow+\infty}\mathbb{E}_{\mathbb{P}}\!\left[U^{-}\!\left(W_{T}^{S}\!\left(\iw,\varphi^{n}\right)\right)\right]
		\leq
		\Xi^{S}\!\left(\varphi\right).
	\end{align*}

	\item
	The desired conclusion can now be reached by reproducing verbatim steps~\ref{item:maxseq} and \ref{item:opt} in the proof of \Cref{th:Ubounded}.%
	\qed
	
\end{enumerate}

\pprangeend{proof:intW}


\paragraph{Proof of \Cref{th:Ureal}}\pprangestart{proof:Ureal}%
Without loss of generality, assume that $U\!\left(0\right)=0$. 
Any strategy in $\mathscr{L}$ is optimal for \eqref{eq:rEU}%
\ if $U\!\left(\cdot\right)$ is constant%
, so suppose otherwise. Then, by concavity of $U\!\left(\cdot\right)$, there exist $C_{1},C_{2}>0$ such that %
\begin{equation}\label{eq:nuage}
	U\!\left(x\right)\leq 
	C_{1}x+C_{2}\ %
	\text{for all }x\leq 0.
\end{equation}
Now %
fix some $S^{\ast}\in\NAset\neq\varnothing$, and %
let $\seq{\varphi^{n}}\subseteq\mathscr{L}$ such that %
	$\inf_{S\in\mathscr{S}}\mathbb{E}_{\mathbb{P}}\!\left[U\!\left(W^{S}_{T}\!\left(\iw,\varphi^{n}\right)\right)\right]>u\!\left(\iw\right)-%
	1/n$ %
for every $n\in\mathbb{N}$ (recall that $u\!\left(\iw\right)>-\infty$). %
As above, we will write $\phi^{n}$ to denote $\hat{\phi}^{S^{*},n}$ in what follows.

The proof consists of several steps%
.
\begin{enumerate}[label=\emph{(\roman*)},wide,labelindent=0pt,labelwidth=!
]

	\item
	Using %
	\begin{align*}
		\mathbb{E}_{\mathbb{P}}\!\left[U^{-}\!\left(W^{S^{\ast}}_{T}\!\left(\iw,\varphi^{n}\right)\right)\right]%
		&\leq U\!\left(+\infty\right)-\inf_{S\in\mathscr{S}}\mathbb{E}_{\mathbb{P}}\!\left[U\!\left(W^{S}_{T}\!\left(\iw,\varphi^{n}\right)\right)\right]\leq U\!\left(+\infty\right)-u\!\left(\iw\right)+\frac{1}{n}%
	\end{align*}
	for all $n\in\mathbb{N}$ and $U\!\left(+\infty\right)<+\infty$%
	, we get %
		$\sup_{n\in\mathbb{N}}\mathbb{E}_{\mathbb{P}}\!\left[U^{-}\!\left(W^{S^{\ast}}_{T}\!\left(\iw,\varphi^{n}\right)\right)\right]<+\infty$. %
	%
	Also, \eqref{eq:nuage} implies %
	\begin{equation*}
		\sup_{n\in\mathbb{N}}\mathbb{E}_{\mathbb{P}}\!\left[\left(W^{S^{\ast}}_{T}\!\left(\iw,\varphi^{n}\right)\right)^{-}\right]\leq\frac{1}{C_{1}}\left(\sup_{n\in\mathbb{N}}\mathbb{E}_{\mathbb{P}}\!\left[U^{-}\!\left(W^{S^{\ast}}_{T}\!\left(\iw,\varphi^{n}\right)\right)\right]+C_{2}\right)<+\infty.
	\end{equation*}
	
	\item\label{item:EMM}
	Since 
	\hyperlink{eq:NAS}{NA($S^{\ast}$)} %
	holds
	, it follows from the Dalang-Morton-Willinger theorem~\citep{dalang1990} that there exists a probability measure $\mathbb{Q}^{\ast}$, equivalent to $\mathbb{P}$, such that $S^{\ast}$ is a martingale under $\mathbb{Q}^{\ast}$, and $\spd^{\ast}\coloneqq
	d\mathbb{Q}^{\ast}/d\mathbb{P}%
	$ is $\mathbb{P}$-a.s.~bounded above by some constant $K^{\ast}>0$%
	. Then, for every $n\in\mathbb{N}$,
	\begin{align}
		\mathbb{E}_{\mathbb{Q}^{\ast}}\!\left[\left(W^{S^{\ast}}_{T}\!\left(\iw,\varphi^{n}\right)\right)^{-}\right]&=\mathbb{E}_{\mathbb{P}}\!\left[\spd^{\ast} \left(W^{S^{\ast}}_{T}\!\left(\iw,\varphi^{n}\right)\right)^{-}\right]%
		\leq K^{\ast} \sup_{n\in\mathbb{N}}\mathbb{E}_{\mathbb{P}}\!\left[\left(W^{S^{\ast}}_{T}\!\left(\iw,\varphi^{n}\right)\right)^{-}\right]
		<+\infty,\label{eq:truemart}
	\end{align}
	so by \citet[Theorem~2]{jacod98} the process $\left\{W^{S^{\ast}}_{t}\!\left(\iw,\varphi^{n}\right)\right\}_{t\in\left\{0,1,\ldots,T\right\}}$ is a $\mathbb{Q}^{\ast}$-martingale. As a consequence,
	\begin{equation*}
		\sup_{n\in\mathbb{N}}\mathbb{E}_{\mathbb{Q}^{\ast}}\!\left[W^{S^{\ast}}_{T}\!\left(\iw,\varphi^{n}\right)^{+}\right]\leq\iw+\sup_{n\in\mathbb{N}}\mathbb{E}_{\mathbb{Q}^{\ast}}\!\left[\left(W^{S^{\ast}}_{T}\!\left(\iw,\varphi^{n}\right)\right)^{-}\right]<+\infty,
	\end{equation*}
	which in turn implies %
		$\sup_{n\in\mathbb{N}}\mathbb{E}_{\mathbb{Q}^{\ast}}\!\left[\left\vert W^{S^{\ast}}_{T}\!\left(\iw,\varphi^{n}\right)\right\vert\right]
		<+\infty$. %
	The fact that $\left\{\left\vert W^{S^{\ast}}_{t}\!\left(\iw,\varphi^{n}\right)\right\vert\right\}_{t\in\left\{0,1,\ldots,T\right\}}$ is a $\mathbb{Q}^{\ast}$-submartingale for each $n\in\mathbb{N}$ leads to %
	\begin{equation}\label{eq:glace}
		\sup_{n\in\mathbb{N}}\sup_{t\in\left\{0,1,\ldots,T\right\}}\mathbb{E}_{\mathbb{Q}^{\ast}}\!\left[\left\vert W^{S^{\ast}}_{t}\!\left(\iw,\varphi^{n}\right)\right\vert\right]\leq\sup_{n\in\mathbb{N}}\mathbb{E}_{\mathbb{Q}^{\ast}}\!\left[\left\vert W^{S^{\ast}}_{T}\!\left(\iw,\varphi^{n}\right)\right\vert\right]<+\infty.
	\end{equation}
	
	\item\label{item:supsup}
	A straightforward adaptation of the arguments from %
	the proof of Lemma~3.11 in %
	\citet
	{rasonyirodriguez2015} shows
	\begin{equation}\label{eq:neige}
		\sup_{n\in\mathbb{N}}\sup_{t\in\left\{1,\ldots,T\right\}}\mathbb{E}_{\mathbb{Q}^{\ast}}\!\left[\frac{\beta^{S^{\ast}}_{t}\left(\kappa^{S^{\ast}}_{t}\right)^{2}}{\spd^{\ast}_{t-1}\mathbb{E}_{\mathbb{P}}\!\left[\nicefrac{1}{\spd^{\ast}_{t}}\left.\right\vert\mathscr{F}_{t-1}\right]}\norm{\varphi^{n}_{t}}\right]<+\infty.
	\end{equation}
	Indeed, if we fix arbitrary $n\in\mathbb{N}$ and %
	$t\in\left\{1,\ldots,T\right\}$, then 
	\eqref{eq:glace} together with
	\begin{equation*}
		\left\vert\innerp{\varphi^{n}_{t}}{\Delta S^{\ast}_{t}}\right\vert%
		=\left\vert W^{S^{\ast}}_{t}\!\left(\iw,\varphi^{n}_{t}\right)-W^{S^{\ast}}_{t-1}\!\left(\iw,\varphi^{n}_{t}\right)\right\vert%
		\leq\left\vert W^{S^{\ast}}_{t}\!\left(\iw,\varphi^{n}_{t}\right)\right\vert+\left\vert W^{S^{\ast}}_{t-1}\!\left(\iw,\varphi^{n}_{t}\right)\right\vert
	\end{equation*}
	yields %
		$\sup_{n\in\mathbb{N}}\sup_{t\in\left\{1,\ldots,T\right\}}\mathbb{E}_{\mathbb{P}}\!\left[\left\vert\innerp{\varphi^{n}_{t}}{\Delta S^{\ast}_{t}}\right\vert\right]<+\infty$. %
	On the other hand, using the tower rule, Bayes' rule, the conditional Cauchy-Schwarz inequality, and
	~\eqref{eq:altNA},
	\begin{align*}
		\lefteqn{%
		\mathbb{E}_{\mathbb{Q}^{\ast}}\!\left[\left\vert\innerp{\varphi^{n}_{t}}{\Delta S^{\ast}_{t}}\right\vert\right]}\\&%
		\geq\mathbb{E}_{\mathbb{Q}^{\ast}}\!\left[\left\vert\innerp{\varphi^{n}_{t}}{\Delta S^{\ast}_{t}}\right\vert\bbOne_{A^{n}_{t}}\right]\geq\mathbb{E}_{\mathbb{Q}^{\ast}}\!\left[\beta^{S^{\ast}}_{t}\norm{\varphi^{n}_{t}}\bbOne_{A^{n}_{t}}\right]
		\geq\mathbb{E}_{\mathbb{Q}^{\ast}}\!\left[\beta^{S^{\ast}}_{t}\norm{\varphi^{n}_{t}}\mathbb{E}_{\mathbb{Q}^{\ast}}\!\left[\left.\bbOne_{A^{n}_{t}}\right\vert\mathscr{F}_{t-1}\right]\right]\\&%
		=\mathbb{E}_{\mathbb{Q}^{\ast}}\!\left[\beta^{S^{\ast}}_{t}\norm{\varphi^{n}_{t}}\frac{1}{\spd^{\ast}_{t-1}}\mathbb{E}_{\mathbb{P}}\!\left[\left.\spd^{\ast}_{t}\bbOne_{A^{n}_{t}}\right\vert\mathscr{F}_{t-1}\right]\right]
		\geq\mathbb{E}_{\mathbb{Q}^{\ast}}\!\left[\beta^{S^{\ast}}_{t}\norm{\varphi^{n}_{t}}\frac{1}{\spd^{\ast}_{t-1}}\mathbb{P}\!\left\{\left.A^{n}_{t}\right\vert\mathscr{F}_{t-1}\right\}^{2}\frac{1}{\mathbb{E}_{\mathbb{P}}\!\left[\left.\nicefrac{1}{\spd^{\ast}_{t}}\right\vert\mathscr{F}_{t-1}\right]}\right]\\&%
		\geq\mathbb{E}_{\mathbb{Q}^{\ast}}\!\left[\beta^{S^{\ast}}_{t}\norm{\varphi^{n}_{t}}\frac{1}{\spd^{\ast}_{t-1}}\left(\kappa^{S^{\ast}}_{t}\right)^{2}\frac{1}{\mathbb{E}_{\mathbb{P}}\!\left[\left.\nicefrac{1}{\spd^{\ast}_{t}}\right\vert\mathscr{F}_{t-1}\right]}\right]%
		,
	\end{align*}
	where $A^{n}_{t}\coloneqq\left\{\innerp{\varphi^{n}_{t}}{\Delta S^{\ast}_{t}}\leq-\beta^{S^{\ast}}_{t}\norm{\varphi^{n}_{t}}\right\}$, and $\spd^{\ast}_{s}\coloneqq\mathbb{E}_{\mathbb{P}}\!\left[\left.\spd^{\ast}\right\vert\mathscr{F}_{s}\right]$ for all $s\in\left\{0,1,\ldots,T\right\}$.

	\item
	Next we claim that, for every $t\in\left\{1,\ldots,T\right\}$, there exists a probability measure $\tilde{\mathbb{Q}}_{t}$, equivalent to $\mathbb{P}$, such that
	\begin{equation*}
		\sup_{n\in\mathbb{N}}\mathbb{E}_{\tilde{\mathbb{Q}}_{t}}\!\left[\norm{\varphi^{n}_{t}}\right]<+\infty.
	\end{equation*}
	The proof is identical to that of e.g.\ \citet[Lemma~3.2]{imkeller2015}, but we reproduce it here for the convenience of the reader. Let $t\in\left\{1,\ldots,T\right\}$ be given, and define $\tilde{\mathbb{Q}}_{t}:~\mathscr{F}_{T}\rightarrow\left[0,1\right]$ as
	\begin{equation*}
		\tilde{\mathbb{Q}}_{t}\!\left(A\right)\coloneqq\frac{\mathbb{E}_{\mathbb{Q}}\!\left[\bbOne_{A}Z_{t}\right]}{\mathbb{E}_{\mathbb{Q}}\!\left[Z_{t}\right]},\quad\text{for all }A\in\mathscr{F}_{T},
	\end{equation*}
	where $Z_{t}\coloneqq\min\!\left\{\beta^{S^{\ast}}_{t}(\kappa^{S^{\ast}}_{t})^{2}/\left(\spd^{\ast}_{t-1}\mathbb{E}_{\mathbb{P}}\!\left[
	1/\spd^{\ast}_{t}\left.\right\vert\mathscr{F}_{t-1}\right]\right),1\right\}>0$ $\mathbb{Q}$-a.s.. %
	Clearly, $\tilde{\mathbb{Q}}_{t}$ is equivalent to $\mathbb{Q}^{\ast}$ (hence, by transitivity, to $\mathbb{P}$). Moreover, for all $n\in\mathbb{N}$,
	\begin{equation*}
		\mathbb{E}_{\tilde{\mathbb{Q}}_{t}}\!\left[\norm{\varphi^{n}_{t}}\right]=\frac{\mathbb{E}_{\mathbb{Q}}\!\left[Z_{t}\norm{\varphi^{n}_{t}}\right]}{\mathbb{E}_{\mathbb{Q}}\!\left[Z_{t}\right]}\leq\frac{1}{\mathbb{E}_{\mathbb{Q}}\!\left[Z_{t}\right]}\sup_{n\in\mathbb{N}}\mathbb{E}_{\mathbb{Q}}\!\left[\frac{\beta^{S^{\ast}}_{t}\left(\kappa^{S^{\ast}}_{t}\right)^{2}}{\spd^{\ast}_{t-1}\mathbb{E}_{\mathbb{P}}\!\left[\nicefrac{1}{\spd^{\ast}_{t}}\left.\right\vert\mathscr{F}_{t-1}\right]}\norm{\varphi^{n}_{t}}\right],
	\end{equation*}
	which combined with \eqref{eq:neige} gives the intended result.

	\item\label{item:Cesaro}
	Since $\sup_{n\in\mathbb{N}}\mathbb{E}_{\tilde{\mathbb{Q}}_{1}}\!\left[\norm{\varphi^{n}_{1}}\right]<+\infty$%
	\ by the previous step%
	, we can apply the Koml\'{o}s theorem to find an $\mathscr{F}_{0}$-measurable $\mathbb{R}^{d}$-valued random variable $\opt_{1}$ with $\mathbb{E}_{\tilde{\mathbb{Q}}_{1}}\!\left[\norm{\opt_{1}}\right]<+\infty$ as well as a subsequence $\seq[k]{\varphi^{n_{1}\!\left(k\right)}_{1}}$ such that all of its subsequences are C\'{e}saro-convergent $\tilde{\mathbb{Q}}_{1}$-a.s.\ 
	to $\opt_{1}$%
	. 
	
	Next, because $\sup_{k\in\mathbb{N}}\mathbb{E}_{\tilde{\mathbb{Q}}_{2}}\!\left[\norm{\varphi^{n_{1}\!\left(k\right)}_{2}}\right]\leq\sup_{n\in\mathbb{N}}\mathbb{E}_{\tilde{\mathbb{Q}}_{2}}\!\left[\norm{\varphi^{n}_{2}}\right]<+\infty$, we can extract a further subsequence $\seq[k]{n_{2}\!\left(k\right)}$ of $\seq[k]{n_{1}\!\left(k\right)}$ such that $\seq[k]{\varphi^{n_{2}\!\left(k\right)}_{2}}$ (with all of its subsequences) C\'{e}saro-converge $\tilde{\mathbb{Q}}_{2}$-a.s.\ to some $\mathscr{F}_{1}$-measurable $\mathbb{R}^{d}$-valued random variable $\opt_{2}$. %
	
	Repeating the same argument $T-2$ more times, we 
	produce %
	$\opt\in\mathscr{L}$ and a subsequence of the original maximising sequence 
	(which, for simplicity, we continue to denote by $\seq{\varphi^{n}}$) such that%
	, for all $t\in\left\{1,\ldots,T\right\}$,
	\begin{equation*}
		\lim_{n\rightarrow+\infty}\frac{1}{n}\sum_{i=1}^{n}\varphi^{i}_{t}=\opt_{t}
	\end{equation*}
	$\tilde{\mathbb{Q}}_{t}$-a.s.\ (whence $\mathbb{P}$-a.s., and consequently also in $\mathbb{P}$-probability).

	\item
	The proof that the functional $\Xi:~\mathscr{L}\rightarrow\mathbb{R}$ defined by
	\begin{equation*}
		\Xi\!\left(\varphi\right)\coloneqq\inf_{S\in\mathscr{S}}\mathbb{E}_{\mathbb{P}}\!\left[U\!\left(W_{T}^{S}\!\left(\iw,\varphi\right)\right)\right],\quad\text{for all }\varphi\in\mathscr{L},
	\end{equation*}
	is upper semicontinuous unfolds exactly as in the proof of \Cref{th:Ubounded}. Hence, 
	\begin{align*}
		\Xi\!\left(\opt\right)\geq\limsup_{n\rightarrow+\infty}\Xi\!\left(\frac{1}{n}\sum_{i=1}^{n}\varphi^{i}\right)\geq\limsup_{n\rightarrow+\infty}\frac{1}{n}\sum_{i=1}^{n}\Xi\!\left(\varphi^{i}\right)=u\!\left(\iw\right),
	\end{align*}
	the second inequality being due to the 
	the assumption that $U\!\left(\cdot\right)$ is concave. %
	This establishes the optimality of $\opt$.%
	\qed

\end{enumerate}

\pprangeend{proof:Ureal}


\paragraph{Proof of \Cref{th:UrealU}}\pprangestart{proof:UrealU}%
We use some ideas originating from \citep{rasonyi2013}, and later reused in \citep{rasonyi2016,rasonyi2017}. We may and will assume that $U\!\left(0\right)=0$; this can be ensured by adding a constant to $U\!\left(\cdot\right)$, which changes neither $\admR{\iw}$ nor the validity of \eqref{maruszja} (though it may change the constant $C$ figuring there). %
The case of constant $U\!\left(\cdot\right)$ is trivial. In all other cases, 
\begin{equation*}
	U\!\left(x\right)\leq C_1 x+C_2\ %
	\text{for all }x\leq 0\tag{\ref{eq:nuage}}
\end{equation*}
holds for some $C_{1},C_{2}>0$. %
We carry out the proof in three steps. %
\begin{enumerate}[label=\emph{(\roman*)},wide,labelindent=0pt,labelwidth=!
]

	\item
	Fix $S\in\mathscr{S}$ and 
	some $c<0$; also, let $\varphi\in\mathscr{L}
	$ %
	with $\mathbb{E}_{\mathbb{P}}\!\left[U^{-}\!\left(W^{S}_{T}\!\left(w_{0},\phi\right)\right)\right]<+\infty$ %
	be given%
	. %
	Define the continuously differentiable, concave, and strictly increasing function $\bar{U}:~\mathbb{R}\rightarrow\mathbb{R}$ as %
	\begin{equation*}
		\bar{U}\!\left(x\right)\coloneqq\left\{
			\begin{array}{ll}
				x,&\text{if }x<0,\\
				2\left(x+1\right)^{1/2}-1,&\text{if }x\geq0
				.%
			\end{array}
		\right.
	\end{equation*}
	By 
	Proposition~7.1 of \citet{rs05} (refer also to Remark~7.4 therein%
	)%
	, there exists a probability measure $\mathbb{Q}\!\left(S\right)\sim\mathbb{P}$ such that $S$ is a $\mathbb{Q}\!\left(S\right)$-martingale, and %
	$d\mathbb{Q}\!\left(S\right)/d\mathbb{P}$ is bounded $\mathbb{P}$-a.s.; furthermore, 
	$d\mathbb{P}/d\mathbb{Q}\!\left(S\right)\in\mathscr{W}$%
	\ (for a similar construction see \citet[Lemma~3.1]{rasonyirodriguez2015})%
	. %
	
	Combining the 
	a.s.\ boundedness of $d\mathbb{Q}\!\left(S\right)/d\mathbb{P}$ %
	with \eqref{eq:nuage} 
	yields $\mathbb{E}_{\mathbb{Q}\!\left(S\right)}\!\left[\left(W^{S}_{T}\!\left(\iw,\varphi\right)\right)^{-}\right]<+\infty$%
	, so by Theorem~2 of \citet{jacod98} the process $\left\{W^{S}_{t}\!\left(\iw,\varphi\right)\right\}_{t\in\left\{0,1,\ldots,T\right\}}$ is a $\mathbb{Q}\!\left(S\right)$-martingale. %
	In particular,
	\begin{equation}\label{zero}
		\mathbb{E}_{\mathbb{Q}\!\left(S\right)}\!\left[W^{S}_{T}\!\left(\iw,\varphi\right)\right]=\iw.
	\end{equation}
	
	Next, choose $\theta>1$ such that $\alpha\theta<1$. Then, %
	\begin{align}
		\mathbb{E}_{\mathbb{P}}\!\left[U^{+}\!\left(W^{S}_{T}\!\left(\iw,\varphi\right)\right)\right]&\leq\mathbb{E}_{\mathbb{P}}\!\left[\left(U^{+}\!\left(W^{S}_{T}\!\left(\iw,\varphi\right)\right)\right)^{\theta}\right]+1\leq C_{3}\left(\mathbb{E}_{\mathbb{P}}\!\left[\left(W^{S}_{T}\!\left(\iw,\varphi\right)^{+}\right)^{\alpha\theta}\right]+1\right)\nonumber\\
		&\leq C_{4}\left(\mathbb{E}_{\mathbb{Q}\!\left(S\right)}\!\left[W^{S}_{T}\!\left(\iw,\varphi\right)^{+}\right]^{\alpha\theta}+1\right)\leq C_{4}\left(\left(\mathbb{E}_{\mathbb{Q}\!\left(S\right)}\!\left[W^{S}_{T}\!\left(\iw,\varphi\right)^{-}\right]+\left\vert\iw\right\vert\right)^{\alpha\theta}+1\right)\nonumber\\
		&\leq C_{5}\left(\mathbb{E}_{\mathbb{P}}\!\left[W^{S}_{T}\!\left(\iw,\varphi\right)^{-}\right]^{\alpha\theta}+1\right)\leq C_{6}\left(\mathbb{E}_{\mathbb{P}}\!\left[U^{-}\!\left(W^{S}_{T}\!\left(\iw,\varphi\right)\right)\right]^{\alpha\theta}+1\right),\label{william}
	\end{align}
	with suitable constants $C_{3},C_{4},C_{5},C_{6}\in\left(0,+\infty\right)$ not depending on the strategy $\varphi$. Here, the first inequality is trivial, while we use \eqref{maruszja} in the second inequality, H\"{o}lder's inequality (with exponent $\nicefrac{1}{\alpha\theta}>1$) and $\nicefrac{d\mathbb{P}}{d\mathbb{Q}\!\left(S\right)}\in\mathscr{W}$ in the third inequality, \eqref{zero} in the fourth inequality, boundedness of $\nicefrac{d\mathbb{Q}\!\left(S\right)}{d\mathbb{P}}$ in the fifth inequality, and \eqref{eq:nuage} in the last inequality. %
	As a consequence,
	\begin{equation*}
		\mathbb{E}_{\mathbb{P}}\!\left[U^{+}\!\left(W^{S}_{T}\!\left(\iw,\varphi\right)\right)\right]<+\infty.
	\end{equation*}
	
	Finally, define 
	\begin{equation*}
		\mathscr{D}_{c}\!\left(S\right)\coloneqq\left\{\varphi\in\mathscr{L}
		:~\mathbb{E}_{\mathbb{P}}\!\left[U\!\left(W^{S}_{T}\!\left(\iw,\varphi\right)\right)\right]\geq c\right\},
	\end{equation*}
	and consider an arbitrary $\varphi\in\mathscr{D}_{c}\!\left(S\right)$. %
	We remark that $\mathscr{D}_{c}\!\left(S\right)$ is a convex subset of %
	\begin{equation*}
		\bar{\mathscr{L}}\!\left(\iw,S\right)\coloneqq\left\{\varphi\in\mathscr{L}:~\mathbb{E}_{\mathbb{P}}\!\left[U^{-}\!\left(W^{S}_{T}\!\left(\iw,\varphi\right)\right)\right]<+\infty\right\} %
	\end{equation*}
	(due to our convention $+\infty-\infty\coloneqq-\infty$). %
	It is elementary to show that, if $x\geq 0$ satisfies $D_{1}\left(1+x^{\alpha\theta}\right)-x\geq D_{2}$ for some $D_{2}<0<D_{1}$, then 
	\begin{equation}\label{elem}
		x\leq\left(2D_{1}\right)^{\frac{1}{1-\alpha\theta}}+2\left(D_{1}-D_{2}\right).
	\end{equation}
	Applying this observation to $D_{1}\coloneqq C_{6}$, $D_{2}\coloneqq c$, and $x\coloneqq\mathbb{E}_{\mathbb{P}}\!\left[U^{-}\!\left(W^{S}_{T}\!\left(\iw,\varphi\right)\right)\right]
	$
	, we get from \eqref{william}, \eqref{elem}, and $\mathbb{E}_{\mathbb{P}}\!\left[U^{+}\!\left(W^{S}_{T}\!\left(\iw,\varphi\right)\right)\right]=\mathbb{E}_{\mathbb{P}}\!\left[U\!\left(W^{S}_{T}\!\left(\iw,\varphi\right)\right)\right]+\mathbb{E}_{\mathbb{P}}\!\left[U^{-}\!\left(W^{S}_{T}\!\left(\iw,\varphi\right)\right)\right]\geq c+\mathbb{E}_{\mathbb{P}}\!\left[U^{-}\!\left(W^{S}_{T}\!\left(\iw,\varphi\right)\right)\right]$ that
	\begin{equation}\label{ll}
		\sup_{\varphi\in\mathscr{D}_{c}\!\left(S\right)}\mathbb{E}_{\mathbb{P}}\!\left[U^{-}\!\left(W^{S}_{T}\!\left(\iw,\varphi\right)\right)\right]<+\infty.
	\end{equation}
	Hence, using \eqref{william} again gives %
	\begin{equation}\label{eq:dlVP}
		\sup_{\varphi\in\mathscr{D}_{c}\!\left(S\right)}\mathbb{E}_{\mathbb{P}}\!\left[\left(U^{+}\!\left(W^{S}_{T}\!\left(\iw,\varphi\right)\right)\right)^{\theta}\right]<+\infty. %
	\end{equation}
	
	\item
	Consider the %
	functional $\Xi^{S}:~
	\bar{\mathscr{L}}\!\left(\iw,S\right)%
	\rightarrow\mathbb{R}$ %
	defined as %
		$\Xi^{S}\!\left(\varphi\right)\coloneqq\mathbb{E}_{\mathbb{P}}\!\left[U\!\left(W^{S}_{T}\!\left(\iw,\varphi\right)\right)\right]$,%
	\ for all $\varphi\in
	\bar{\mathscr{L}}\!\left(\iw,S\right)%
	$. %
	Similarly to the previous proofs, we show upper semicontinuity of	$\Xi^{S}$ on $\mathscr{D}_{c}\!\left(S\right)$
	. If $\mathscr{D}_{c}\!\left(S\right)$ is empty, then there is nothing to prove. Otherwise, let $\seq{\psi^{n}}\subseteq\mathscr{D}_{c}\!\left(S\right)$ converge to some $\psi$ (in the topology of $\mathscr{L}$). 
	By de la Vall\'{e}e-Poussin theorem, \eqref{eq:dlVP} implies the uniform integrability of the family $\left\{U^{+}\!\left(W^{S}_{T}\!\left(\iw,\psi^{n}\right)\right)\right\}_{n\in\mathbb{N}}$, so %
\begin{equation*}
	\Xi^{S}\!\left(\psi\right)\geq\limsup_{n\rightarrow\infty}\Xi^{S}\!\left(\psi^{n}\right)\geq c%
	,
\end{equation*}
which entails not only upper semicontinuity of $\Xi^{S}\!\left(\cdot\right)$, but $\psi\in\mathscr{D}_{c}\!\left(S\right)$ as well. In other words,
we have also shown that $\mathscr{D}_{c}\!\left(S\right)$ is closed in $\mathscr{L}$. Then it is easy to see that 
$\Xi:~\admR{\iw}\rightarrow\mathbb{R}$ given by %
\begin{equation*}
	\Xi(\varphi)\coloneqq\inf_{S\in\mathscr{S}}\Xi^{S}\!\left(\varphi\right),\quad\text{for all }\varphi\in\admR{\iw},
\end{equation*}
is upper semicontinuous, too, on $\cap_{S\in\mathscr{S}}\mathscr{D}_{c}\!\left(S\right)$.

	\item
	Now we turn to showing existence of an optimiser. Take a sequence $\seq{\phi^{n}}\subseteq\admR{\iw}$ such that $\Xi\!\left(\phi^{n}\right)\rightarrow\sup_{\varphi\in\admR{\iw}}\Xi\!\left(\varphi\right)%
	$ as $n\rightarrow\infty$. There is $c<0$ such that $\phi^{n}\in\cap_{S\in\mathscr{S}}\mathscr{D}_{c}\!\left(S\right)$ for all $n\in\mathbb{N}$ large enough %
	(otherwise we could extract a subsequence $\seq[k]{\phi^{n_{k}}}$ with $\Xi\!\left(\phi^{n_{k}}\right)<-k$ for all $k\in\mathbb{N}$, leading to the contradiction $u\!\left(\iw\right)=-\infty$). Fix an arbitrary $S^{\ast}\in\mathscr{S}$. Recalling~\eqref{eq:nuage}, \eqref{ll}, and the boundedness of $\nicefrac{d\mathbb{Q}\!\left(S^{\ast}\right)}{d\mathbb{P}}$, 
	\begin{equation*}
		\sup_{n\in\mathbb{N}}\mathbb{E}_{\mathbb{Q}(S^{\ast})}[(W^{S^{\ast}}_T(\iw,\phi^n))^{-}]<+\infty,
	\end{equation*}
	which also implies, by the $\mathbb{Q}\!\left(S^{\ast}\right)$-martingale property of $\left\{W^{S^{\ast}}_{t}\!\left(\iw,\phi^{n}\right)\right\}_{t\in\left\{0,1,\ldots,T\right\}}$ for each $n\in\mathbb{N}$, that
	\begin{equation*}
		\sup_{n\in\mathbb{N}}\mathbb{E}_{\mathbb{Q}\!\left(S^{\ast}\right)}\!\left[\left\vert W^{S^{\ast}}_{T}\!\left(\iw,\phi^{n}\right)\right\vert\right]<+\infty.
	\end{equation*}
%
%
From this point onwards, we can follow %
\emph{mutatis mutandis} the steps~\ref{item:supsup} to \ref{item:Cesaro} in %
the proof of \Cref{th:Ureal} and %
obtain an optimal strategy $\opt\in\cap_{S\in\mathscr{S}}\mathscr{D}_{c}\!\left(S\right)\subseteq\admR{\iw}$.%
\qed

\end{enumerate}

\pprangeend{proof:UrealU}


\bibliographystyle
{chicago}
\bibliography{C:/Users/Andrea/Documents/AndreaDocuments/biblio}


\end{document}